\documentclass{article}

\usepackage{arxiv}
\usepackage{amsmath,bm}
\newcommand{\comment}[1]{}
\usepackage[utf8]{inputenc} % allow utf-8 input
\usepackage[T1]{fontenc}    % use 8-bit T1 fonts
\usepackage[colorlinks=true, linkcolor=blue, urlcolor=blue, citecolor=blue]{hyperref}
\usepackage{url}            % simple URL typesetting
\usepackage{booktabs}       % professional-quality tables
\usepackage{amsfonts}       % blackboard math symbols
\usepackage{nicefrac}       % compact symbols for 1/2, etc.
\usepackage{microtype}      % microtypography
\usepackage{lipsum}		% Can be removed after putting your text content
\usepackage[most]{tcolorbox}
\usepackage{graphicx}
\usepackage{natbib}
\usepackage{doi}
\usepackage[noabbrev,capitalise]{cleveref}
\crefformat{section}{#2\S~#1#3}
\Crefformat{section}{#2\S~#1#3}
\crefformat{subsection}{#2\S~#1#3}
\Crefformat{subsection}{#2\S~#1#3}

\title{From Modules to Movement: Deconstructing the Modular Architecture of the Motor System}

\author{{Alessandro Salatiello${^{1,2}}$}\\
	$^1$University of Tübingen \\
    $^2$International Max Planck Research School for Intelligent Systems\\
	\texttt{alessandro.salatiello@uni-tuebingen.de} \\
	}
\date{}

\renewcommand{\shorttitle}{Deconstructing the Modular Architecture of the Motor System}

\hypersetup{
pdftitle={A template for the arxiv style},
pdfsubject={q-bio.NC, q-bio.QM},
pdfauthor={Alessandro Salatiello},
pdfkeywords={xxx, yyy, zzz},
}

\begin{document}
\maketitle

\begin{abstract}
Coordinating multi-articulated bodies to generate purposeful movement is a formidable computational challenge. Yet the human motor system performs this task robustly in dynamic, uncertain environments, despite noisy and delayed feedback, slow actuators, and strict energetic constraints. A central question is what organizational principles underlie this efficiency. One widely recognized principle of neural organization is modularity, which enables complex problems to be decomposed into simpler subproblems that specialized modules are optimized to solve. In this review, we argue that modularity is a fundamental organizing principle of the motor system. We first summarize evidence for brain modularity, ranging from classical lesion studies to contemporary graph-theoretical analyses. We next discuss the main factors underlying the emergence and evolutionary selection of modular architectures, highlighting the computational advantages they provide. We then review the major neuroanatomical modules that structure current descriptions of the motor system and compare three prominent computational frameworks of motor control—optimal feedback control theory, muscle synergy theory, and dynamical systems approaches—showing that all implicitly or explicitly rely on specialized computational modules. We conclude by contrasting the key strengths and limitations of existing frameworks and by proposing promising directions toward more comprehensive theories.
\end{abstract}

% Start the table of contents on a new page
\tableofcontents
\newpage

\section{Introduction}
\label{sec:intro}
Controlling multi-articulated bodies in an ever-changing environment is tremendously hard. And yet, brains excel at this task despite relying on only 20W of power, slow actuators, and noisy sensory feedback. One of the established features that allows brains to be so effective is their modular organization. This property, common across several natural complex systems, facilitates solving complex problems by decomposing them into sub-problems that specialized modules can tackle more easily \citep{salatiello2025modularity}. Additionally, when these modules can be arbitrarily reused and recombined in new contexts, they enable the solution of an exponentially larger set of complex problems. 

What makes modular designs so efficient? \citet{baldwin1999design} investigated this question and identified six fundamental properties underlying their effectiveness. Specifically, they introduced six {\it fundamental operators} that modular architectures inherently support. These operators facilitate not only {\it splitting} a problem into manageable subproblems, but also {\it substituting} components with improved alternatives; {\it augmenting} systems with new functionality, {\it excluding} unnecessary components; {\it porting} components across different architectures; and {\it inverting}, which consolidates specific solutions into reusable components. These properties have contributed to establishing modularity as a central design principle in engineering, where robustness and reusability are paramount.

Modular architectures can thus facilitate the development of increasingly sophisticated yet robust systems capable of adapting to and thriving in dynamic environments — much like biological organisms do. Accordingly, modularity is often proposed as a key feature underlying the remarkable cognitive and sensorimotor capabilities of animals. Specifically, in the case of the motor system, a variety of modular frameworks have been developed to explain its organizational structure. However, these frameworks differ in how they define the fundamental modules, reflecting diverse methodological and theoretical perspectives, which are often hard to reconcile. While previous works have reviewed individual frameworks, they have typically done so in isolation, limiting our ability to assess their relationships and broader implications. In this review, we aim to address this gap by synthesizing these perspectives, identifying their areas of convergence and divergence, and clarifying how they collectively inform our understanding of modularity in the motor system.

In what follows, we first present neuropsychological, neuroanatomical, and graph-theoretical evidence supporting brain modularity (\cref{sec:evidence}), and then discuss why the brain might be modular, drawing on insights from complex systems theory, evolutionary biology, and graph theory (\cref{sec:why_modular}). We then turn to the motor system and examine the various ways it has been modeled as modular, drawing on neuroanatomical (\cref{sec:mod_motor_system:anatomy}), engineering (\cref{sec:mod_motor_system:engineering}), dimensionality reduction (\cref{sec:mod_motor_system:data_driven_output}), and dynamical systems (\cref{sec:mod_motor_system:data_driven_brain}) perspectives. Specifically, we begin with neuroanatomical descriptions based on the organization of cortical and subcortical areas such as the motor cortex, the basal ganglia, the cerebellum, the brainstem, and the spinal cord (\cref{sec:mod_motor_system:anatomy}). Next, we examine formalizations that approach the problem of controlling the body from an engineering perspective, emphasizing the need for dedicated components for robust state-dependent control (\cref{sec:mod_motor_system:engineering}). We next examine theories positing that the motor system leverages modular components that encode reusable coordination patterns to mitigate musculoskeletal redundancy and reduce the dimensionality of the muscle control signals  (\cref{sec:mod_motor_system:data_driven_output}). Finally, we review frameworks that explain motor system organization in terms of structured computations unfolding within distinct subspaces of the neural population activity (\cref{sec:mod_motor_system:data_driven_brain}). We conclude by summarizing the main insights and limitations inherent in these frameworks, and by outlining open questions for future work (\cref{sec:conclusion}).

\section{Evidence of brain modularity} %Is the brain modular
\label{sec:evidence}
Compelling evidence for brain modularity comes from multiple lines of research. In the following sections, we highlight three particularly influential approaches. We first review neuropsychological findings from classical and modern lesion studies demonstrating that damage to circumscribed brain regions produces selective functional deficits \cref{sec:evidence:lesion}. We then turn to neuroanatomical parcellation studies, which reveal spatially discrete cortical areas with distinctive structural and functional properties \cref{sec:evidence:parcellation}. Finally, we examine graph-theoretical analyses showing that brain networks exhibit modular community structure alongside global integrative features such as small-worldness and rich-club organization \cref{sec:evidence:graph-theory}.

\subsection{Lesion studies}
\label{sec:evidence:lesion}
One of the earliest sources of evidence for brain modularity comes from classical lesion studies, which showed that damage to specific brain regions often resulted in well-defined deficits, while leaving other cognitive, perceptual, and motor functions largely intact \citep{vaidya2019lesion}. Below, we highlight three of the most influential cases. Phineas Gage suffered a traumatic lesion to the ventromedial prefrontal cortex, which selectively disrupted his decision-making and emotion-processing abilities \citep{harlow1848passage,damasio1994return}.
Louis Victor Leborgne and Lazare Lelong had lesions in the posterior inferior frontal gyrus (a region that came to be known as Broca's area), which selectively affected their ability to speak \citep{broca1861remarques,broca1861perte,dronkers2007paul}.
Henry Gustave Molaison (H.M.) underwent a bilateral medial temporal lobectomy to treat his epilepsy and lost his ability to form new long-term memories of facts and events, while maintaining the ability to recall pre-surgery memories, use working memory, and form new motor memories \citep{scoville1957loss, squire2009legacy}. Although subsequent research \citep{damasio1994return,dronkers2007paul,thiebaut2015phineas} has revealed that both the structural damage and functional impairments in these cases were more extensive than initially believed\footnote{These studies also established that the affected regions were involved in more functions than initially assumed, and that the observed symptoms depended on broader networks.}, these studies still provide foundational evidence for the modular organization of the brain. 

The historical cases discussed above involved systems other than the motor system. However, similar principles emerge from the analysis of motor system lesions: damage to specific cortical and subcortical regions in the motor system consistently leads to distinct and well-characterized motor deficits \citep{shadmehr2008computational}. For example, damage to the cerebellum is commonly associated with impaired multi-joint coordination, difficulty predicting the sensory consequences of motor commands, and reduced ability to adapt those commands to changes in the mapping between motor commands and sensory feedback \citep{shadmehr2008computational}. Similarly, hyperactivation of the indirect pathway of the basal ganglia --- caused by the degeneration of dopaminergic neurons in the substantia nigra pars compacta in Parkinson's disease --- leads to excessive inhibition of the thalamus and thus insufficient excitation of the motor cortex. This cascade results in characteristic hypokinetic symptoms such as difficulty initiating movement and bradykinesia \cite{mcgregor2019circuit,kandel2021principles}. On the contrary, hypoactivation of the indirect pathway of the basal ganglia --- caused by the degeneration of dopamine-sensitive neurons in the striatum in Huntiton's disease --- leads to reduced inhibition of the thalamus and thus excessive excitation of the motor cortex. This results in hyperkinetic symptoms such as involuntary and uncontrolled movements \citep{mcgregor2019circuit,kandel2021principles}. Finally, lesions to the primary motor cortex often result in marked muscle weakness on the contralateral side, impaired dexterity, and abnormally increased muscle co-activation, reflecting the primary motor cortex's crucial role in fine motor control \citep{shadmehr2008computational,ting2015neuromechanical}.

In conclusion, although lesion studies have important limitations --- most notably the difficulty of ruling out compensatory contributions of intact brain regions --- they remain a valuable starting point for identifying the functions of specific brain regions \citep{rorden2004using}. Most importantly for our purposes, they offer compelling evidence for brain modularity and offer insight into the organization and roles of its constituent modules\citep{shadmehr2008computational}.

\subsection{Parcellation studies}
\label{sec:evidence:parcellation}

Another influential line of evidence supporting brain modularity comes from analyses of the functional and histological properties of the cerebral cortex \cite{petersen2024principles}. These investigations consistently reveal that the cortex is tiled by spatially discrete patches of neurons exhibiting characteristic structural and functional properties.

A landmark functional study is the pioneering work of \cite{fritsch1870uber}, which demonstrated the existence of a well-delineated area of the cortex --- the motor cortex --- with the special property that its electrical stimulation elicits muscle twitches, with specific cortical loci preferentially activating particular muscle groups. Subsequent studies \citep{ferrier1873experimental} demonstrated that longer electrical stimulation evokes more complex, multi-joint movements \citep{taylor2003twitches}. Later, more detailed analyses of the functional properties of these regions led to finer maps and clarified that cortical areas that preferentially activate different muscle groups can overlap \citep{penfield1937somatic}. Further work established that the elicited movements are often behaviorally meaningful, converging to site-specific final postures \citep{graziano2002complex}. These findings support the existence of functional modules representing ethologically relevant, multi-joint actions, such as hand-to-mouth or defensive movements \cite{graziano2002complex}.

From the histological perspective, very influential contributions include the {\it cytoarchitectonic} cortical maps developed by \cite{brodmann1909vergleichende} and \cite{von1925cytoarchitektonik} --- based on the inter-area differences in the density, shape, size, and laminar distribution of cell bodies \cite{amunts2015architectonic} --- and the {\it myeloarchitectonic} cortical maps developed by \cite{vogt1919allgemeine} --- based on density, orientation, and laminar distribution of myelinated axons \cite{van2018parcellating}. These classical studies identified between 43 \cite{brodmann1909vergleichende} and 185 \cite{vogt1919allgemeine} distinct cortical areas \cite{amunts2015architectonic}.

Over the past two decades, interest in identifying the fundamental cortical processing units, or {\it parcels}, has resurged, fueled by advances in neuroimaging techniques such as MRI and fMRI \cite{eickhoff2018topographic} and by the development of observer-independent, semi-automated algorithms. These innovations have enabled in vivo estimation of structural and functional cortical properties across multiple brains. In such {\it parcellation studies}, a cortical area is typically defined as a spatially contiguous region that differs from its neighbors in one or more of the following four fundamental properties: cyto-, myelo-, or receptor-architecture (structure), inter-area connectivity, function, and topographic organization \cite{van2018parcellating,petersen2024principles}.

While most parcellation studies rely on a single imaging modality, which limits their ability estimate all fundamental properties reliably (e.g., \cite{desikan2006automated,gordon2016generation,schaefer2018local}), others integrate multiple modalities and can thus leverage several properties simultaneously to delineate the parcels (e.g., \cite{fan2016human,glasser2016multi,huang2022extended}). For example, \cite{glasser2016multi} designed a semi-automatic multimodal approach that led to the robust identification of 180 areas. Specifically, they leveraged MRI-based measures of myelin content and cortical thickness to identify structural boundaries, resting-state fMRI to estimate functional connectivity and topographic organization, and task-based fMRI from seven behavioral paradigms to define functionally distinct regions. 

In summary, despite variability in methodological approaches and in the estimated number of cortical areas, converging evidence supports the existence of spatially discrete, homogeneous \textit{parcels}—modules with distinctive structural and functional properties—within the cortex.

\subsection{Graph-theoretical studies}
\label{sec:evidence:graph-theory}

Another strong line of evidence for brain modularity comes from graph-theoretical analyses of structural and functional connectivity in both animal models and humans \citep{meunier2010modular,sporns2016modular}. In humans, structural graphs are typically constructed by treating cortical areas as nodes and structural connections—often estimated with MRI-based diffusion tensor imaging—as edges. Conversely, functional graphs are typically built using functional connections --- estimated by measuring the correlations between functional MRI (fMRI) time series --- as edges. Once built, these graphs can be studied using tools from graph theory \citep{bullmore2009complex}.

A key insight from this approach is that both structural and functional brain graphs are {\it small-world} networks \citep{watts1998collective,bassett2006small,bassett2017small}. Such networks are characterized by high clustering (above-chance presence of triangular motifs) and short characteristic path lengths (low average pairwise shortest-path distance). Small-worldness, a property shared by many complex systems, is thought to support local specialization while maintaining efficient global communication \citep{bassett2006small,bassett2017small}.

Small-worldness captures the {\it local} tendency of nodes to form triangular clusters and their {\it global} tendency to be connected through topological shortcuts that facilitate long-range communication \citep{meunier2010modular}. However, analyses at intermediate scales have revealed an additional level of organization. Specifically, nodes tend to cluster into {\it communities} --- or {\it modules} --- of densely interconnected nodes that are sparsely connected to nodes of other communities \citep{sporns2016modular}. Furthermore, these communities are typically linked by {\it connector hubs} --- nodes with a disproportionately large number of connections to nodes of other modules. In turn, connector hubs tend to be densely interconnected and thus form the so-called {\it rich clubs}, which are believed to support efficient inter-module communication and information integration \citep{van2011rich}. For example, \citet{hagmann2008mapping} identified six modules from MRI-derived structural connectivity data. Four were bilaterally symmetric and located in the frontal and temporo-parietal cortices, while two were located medially on the precuneus and posterior cingulate cortex. Within modules, communication was facilitated by provincial hubs, especially abundant within the four most posterior modules. Integration across modules was instead mediated by a rich club of connector hubs, concentrated medially within the two medial modules and the medial regions of the frontal modules. 

Together, these findings indicate that modular organization is also supported by graph-theoretical analyses of brain networks. From a broader perspective, such analyses have shown that properties like small-worldness, modularity, and rich-club organization appear universal across diverse brain networks. They are consistently observed in systems ranging from electron microscopy–derived structural connectomes in C. elegans \citep{watts1998collective} to fMRI-based functional networks in humans \citep{salvador2005neurophysiological}, and they therefore represent a key principle of brain organization \citep{bullmore2009complex,van2016comparative}.

\section{Why are brains modular?} %Why might this be the case
\label{sec:why_modular}
But how did such a modular organization develop in the brain? Phylogenetic studies suggest that the current architecture might be the result of the gradual duplication and specialization of ancestral pools of neural modules \citet{cisek2019resynthesizing,chakraborty2015brain,feenders2008molecular}. Such a process provided an effective means for surviving in a world that is dynamic, yet tends to change in ways that preserve much of its underlying structure.

Consistent with this evolutionary trajectory, modules that serve broad, fundamental functions tend to be conserved across species, while new functional demands are often met through the emergence of novel modules. For example, the regulation of internal states and the capacity to approach beneficial stimuli while avoiding harmful ones are fundamental requirements across virtually all animals. Accordingly, many homeostasis-regulating nuclei of the hypothalamus and locomotor-controlling nuclei in the brainstem and spinal cord are highly conserved across vertebrates, including species as evolutionarily distant as mammals and lampreys \citep{cisek2022neuroscience}. In fact, virtually all the brainstem and spinal cord modules responsible for vital motor programs, such as respiration, swallowing, and eye movements, are highly conserved across vertebrates, along with the basal ganglia circuits mediating their activation \citep{grillner2016basal,grillner2023brain}.

However, what qualifies as a beneficial or harmful stimulus can vary greatly depending on ecological context, and adapting to these differences may require anything from minor adjustments in existing neural networks to more substantial rewiring. For instance, transitioning from light-approaching to light-avoiding behaviors may be implemented by simple modifications in stimulus-response associations, without requiring new circuitry. 
In contrast, distinguishing benevolent from malevolent peers in primates placed selective pressure on the evolution of new modules that provided the capacity to recognize familiar individuals and interpret social signals \citep{darwin1993expression,nelson2001development,leopold2010comparative}. Notable examples of such modules include the fusiform face area (FFA) and the superior temporal sulcus (STS), which are established specialized regions for face and emotion processing \citep{tsao2008comparing,hesse2020macaque}. In both cases, modular architectures provide the structural flexibility to maintain core functions while incorporating new ones, enabling more effective adaptation to novel ecological and social challenges. 

In the following sections, we review influential work that has sought to formalize and extend these ideas, with particular emphasis on evolutionary perspectives. We first examine established theories concerning the emergence of modularity \cref{sec:why_modular:emergence}. We then turn to explanations for why modularity may have been selected for by natural evolution, considering both direct and indirect contributions to fitness. 

\subsection{Emergence of modularity}
\label{sec:why_modular:emergence}
Several theoretical perspectives have been proposed to explain the emergence of modularity; in this section, we will focus on four particularly influential ones. Some perspectives view modules as stable configurations that arise spontaneously in complex systems \cref{ss:stability}. Others emphasize biological mechanisms such as gene duplication and differentiation as the key underlying factor driving the emergence of modules \cref{ss:duplication}. Finally, other perspectives argue that modularity evolves as an adaptive response to dynamic environments \cref{ss:dynamic_env}, or as a solution to energetic constraints \cref{ss:energy}.

\subsubsection{Stability of random aggregations}
\label{ss:stability}
Brains are not the only systems widely regarded as modular; most natural complex systems—typically defined as assemblies of interacting components with collective dynamics giving rise to emergent behavior—also exhibit modular organization. This ubiquity points to the operation of a general principle \citep{callebaut2005modularity}. \citet{simon2012architecture} put forth such a principle, proposing that modularity in complex natural systems arises because modules act as stable intermediate units that provide greater robustness to perturbations. In non-modular systems, perturbations can cascade throughout the system and potentially dissolve its structure, whereas in modular systems their effects are largely confined to the affected module. Moreover, the aggregation of modules into higher-order super-modules confers the same advantages at larger scales, giving rise to increasingly sophisticated, hierarchically modular systems \citep{simon2012architecture,sales2007extracting}.

\subsubsection{Gene duplication and differentiation}
\label{ss:duplication}
Modularity is observed across nearly all levels of biological organization \citep{wagner2007road}. While many theories attribute its emergence to environmental pressures, others suggest it may arise as a byproduct of environment-independent, lower-level fundamental mechanisms not directly selected for \citep{wagner2007road}. One such mechanism is gene duplication and differentiation \citep{ohno2013evolution,lynch2000probability}, which has been proposed as a driver of modularity across several organizational scales, including in protein interaction networks \citep{sole2002model}. Supporting this view, simulations show that artificial neural networks incorporating gene duplication and differentiation become modular \citep{hallinan2004gene}, with resulting modules exhibiting functional specialization \citep{calabretta2000duplication}. This mechanism is also thought to operate at higher levels, contributing to brain pathway duplication \citep{feenders2008molecular}, a process implicated in the evolution of novel brain functions \citep{chakraborty2015brain}. Thus, gene duplication and differentiation may drive the emergence of modularity across multiple biological scales.  

\subsubsection{Adaptation to dynamic environments}
\label{ss:dynamic_env}
Biological organisms must survive in dynamic environments that continually impose new challenges. A seminal study by Lipson and colleagues \citep{lipson2002origin} simulated such conditions and showed that environmental variability alone was sufficient to induce modularity in artificial neural networks, with the degree of modular separation increasing as the rate of change accelerated. Subsequent work \citep{kashtan2005spontaneous} demonstrated that when environments change in a modular fashion (i.e., with fixed subgoals), the resulting modular structure becomes even more pronounced. Critically, this emerging modular organization was shown to markedly accelerate adaptation to the environment, especially for challenging tasks \citep{kashtan2007varying}. The underlying principle is that switching between environments with shared subgoals fosters the emergence of modules specialized for those subgoals, thereby facilitating adaptation. By contrast, in static environments, such subgoal-specific modules do not develop, and solutions that lack task decomposition must be learned instead. Collectively, these studies provide compelling evidence that environmental variability promotes modularity, and that modularity in turn enhances adaptability—a fundamental requirement for survival \citep{bateson2017adaptability}.

The relationship between task demands and emerging network organization has been explored not only in studies of evolution but also in work aimed at uncovering fundamental learning principles of natural and artificial intelligence \citep{yang2019task,dobs2022brain,johnston2023abstract,vafidis2024disentangling,gu2025task}. These studies demonstrate that networks trained on multiple tasks with shared substructure develop functional modularity \citep{yang2019task,dobs2022brain,gu2025task} and learn abstract task representations that support out-of-distribution generalization \citep{johnston2023abstract,vafidis2024disentangling}. Thus, survival in a dynamic world requires the ability to learn to solve multiple tasks quickly. This environmental pressure promotes the emergence of modular architectures, which not only speed up learning but also facilitate generalization to novel tasks.

\subsubsection{Energetic constraints}
\label{ss:energy}
Biological organisms must survive in environments that are not only dynamic but also energy limited, and thus they face strong selective pressures to optimize energy expenditure. This constraint is likely to place strong selective pressure on the organization of the brain, one of the most metabolically demanding organs in the human body, which consumes roughly 20\% of total energy resources \citep{herculano2011scaling}. A large fraction of this energy is devoted to maintaining the electrochemical gradients across neuronal membranes, which is necessary for the generation and transmission of of action potentials \citep{niven2008energy}. Because this energy cost scales with axon length \citep{bullmore2012economy}, minimizing axonal wiring becomes energetically efficient\footnote{Furthermore, shorter wiring not only reduces metabolic cost but also conserves volume within the limited space of the skull \citep{bullmore2012economy}.}. Consistent with this principle, brain networks exhibit a tendency to prefer local over long-distance connections \citep{hellwig2000quantitative}. This preference contributes to the emergence of modular structure, since spatial proximity often coincides with topological proximity in brain networks \citep{meunier2009hierarchical}.

A growing body of computational work has further clarified the relationship between wiring minimization and modularity \citep{jacobs1992computational,clune2013evolutionary,ellefsen2015neural,liu2023seeing,achterberg2023spatially}. Networks optimized to maximize performance while minimizing wiring length consistently develop modular architectures, which additionally exhibit a range of computational advantages (discussed in \cref{sec:why_modular:direct}).

Together, these findings provide converging evidence that wiring minimization promotes the emergence of modular organization. Nevertheless, wiring minimization alone cannot fully account for the topology of brain connectomes. Brains also contain energetically costly features, including long-range connections and tightly connected connector hubs. These features, while metabolically expensive, substantially enhance network efficiency: long-range connections reduce average path length (cf. \cref{sec:evidence:graph-theory}), and connector hubs facilitate intermodular communication and integrative processing. In this way, brains appear to balance energetic constraints against functional demands, accepting higher metabolic costs when they yield significant gains in performance.

\subsection{Selection for modularity}
\label{sec:why_modular:selection}
In \cref{sec:why_modular:emergence}, we reviewed leading theories on how modularity arises in biological networks. For such networks to be favored by natural selection, their modular organization must ultimately confer a fitness advantage. In this section, we examine established benefits attributed to modularity, which can be both direct and indirect. Directly, modularity enhances computational efficiency and functional flexibility \cref{sec:why_modular:direct}. Indirectly, it underpins broader principles such as robustness, evolvability, and adaptability \cref{sec:why_modular:indirect}.

\subsubsection{Computational benefits}
\label{sec:why_modular:direct}
Networks trained to jointly optimize performance and wiring costs not only develop modular structure (c.f. \cref{ss:energy}) but also exhibit clear computational advantages. Such networks decompose tasks into subtasks with decoupled internal representations and dynamics \citep{jacobs1992computational}, acquire new skills more rapidly \citep{clune2013evolutionary} and with greater resistance to catastrophic forgetting \citep{ellefsen2015neural}. Additionally, they reuse features in a compositional manner \citep{liu2023seeing}, display low mean activation and brain-like topological features \citep{achterberg2023spatially}. As discussed in \cref{ss:dynamic_env}, networks trained on multiple tasks also develop modular architectures. Such networks also display several advantageous properties, including the emergence of abstract task representations \citep{johnston2023abstract} and functionally specialized modules \citep{yang2019task,dobs2022brain,gu2025task}, which in turn facilitate out-of-distribution generalization \citep{vafidis2024disentangling}.

A complementary line of work has investigated modularity by imposing it a priori or enforcing it through bespoke algorithms. These studies consistently show that modular networks tend to be more data-efficient and achieve higher asymptotic performance than monolithic counterparts \citep{jacobs1991adaptive,andreas2016neural,kirsch2018modular,khona2023winning}. This advantage is generally attributed to their ability to generalize compositionally: by learning reusable task components that can be flexibly recombined, modular networks avoid the need for exponentially large training sets to handle novel inputs \citep{boopathy2024breaking}. However, these benefits do not always arise spontaneously. In practice, end-to-end training can yield modules that fail to align with the underlying task structure \citep{csordas2020neural,mittal2022modular,jarvis2024specialization}. Specialized learning algorithms \citep{boopathy2024breaking} or constraints on computational resources \citep{bena2025dynamics} may therefore be required to encourage alignment and unlock the full computational potential of modular architectures.

Several studies have also examined the dynamics of networks with modular architectures. These works show that networks composed of {\it sparsely} connected modules exhibit rich dynamics characterized by multistability and metastability \citep{sporns2000theoretical,palma2025balance}, as well as the emergence of multiple timescales \citep{arenas2006synchronization,pan2009modularity}. Intuitively, when modules are strongly interconnected, their activity becomes synchronized—similar to random networks—yielding relatively homogeneous, flat dynamics. Conversely, when modules are completely disconnected, no meaningful interactions can occur, leading again to trivial dynamics. In contrast, sparse intermodular connectivity strikes a balance, enabling both specialized local processing and integrative global coordination \citep{sporns2000theoretical,palma2025balance}.

\subsubsection{Indirect advantages}
\label{sec:why_modular:indirect}
As discussed in \cref{sec:why_modular:direct}, modularity confers computational benefits—such as faster and more efficient learning—that are likely to directly improve fitness and thereby increase the likelihood of evolutionary selection. Beyond these immediate advantages, modularity is also thought to indirectly enhance fitness by promoting the emergence of properties that do not necessarily benefit the individual organism in the short term, but confer advantages across evolutionary timescales \citep{wagner2007road,lorenz2011emergence}.

A prominent example is the ability to accommodate {\it lateral, or horizontal, gene transfer}—the non-parental exchange of genetic material between organisms—which has played a major role in the evolution of both prokaryotic and eukaryotic genomes \citep{keeling2008horizontal}. When successfully integrated, transferred genes can endow recipient organisms with novel capabilities (e.g., enabling the exploitation of a previously inaccessible nutrient), thereby conferring a substantial fitness advantage. \cite{rainey2004evolution} proposed that modular regulatory architectures—where regulatory genes preferentially control functionally related structural genes—are particularly well-suited to harnessing the benefits of horizontal transfer while minimizing deleterious pleiotropic effects. In modular networks, an acquired gene can be efficiently regulated by a dedicated controller without disrupting unrelated pathways, thereby preserving existing functions while integrating new ones. By contrast, in non-modular networks, regulatory entanglement makes such integration less feasible, as optimizing the expression of new DNA may inadvertently perturb existing pathways. This capacity to assimilate foreign genetic material may help explain why modularity is indirectly favored by natural selection.

Another fitness-enhancing property associated with modularity is {\it robustness} (c.f. \cref{ss:stability}). At the phenotypic level, robustness enables perturbations to remain contained within individual modules, preventing their spread across the system and conferring a direct fitness advantage—for example, in modular brain networks. At the genetic level, robustness of regulatory pathways makes an indirect contribution to fitness, with benefits that manifest over evolutionary timescales. In modular architectures, deleterious mutations confined to a single functional module are less likely to compromise the viability of the whole organism, allowing non-lethal mutations to accumulate and, across generations, potentially give rise to beneficial innovations. Conversely, advantageous mutations restricted to a single module are more likely to improve fitness, as they are less prone to produce harmful pleiotropic effects elsewhere in the network. Moreover, because modules often map directly onto specific phenotypic traits, beneficial genetic changes can produce distinct phenotypic variations that are available to selection.

Genetic robustness translates into increased {\it evolvability}—the capacity of organisms to generate heritable, adaptive phenotypic variation \citep{kirschner1998evolvability,gerhart2007theory,payne2019causes}—for which modularity is thought to be a key driver. Notably, modularity has been linked to both adaptability \citep{kashtan2007varying} and evolvability \citep{kirschner1998evolvability}, traits that are themselves believed to be interconnected: enhanced adaptability allows organisms to survive under changing conditions, thereby providing the time and opportunity for evolutionary processes to unfold \citep{bateson2017adaptability}.

\section{Modularity in the motor system}
\label{sec:mod_motor_system}
Multiple lines of evidence support the view that the motor system is organized in a modular fashion. In this section, we review complementary perspectives on motor modularity, drawing on neuroanatomical and functional considerations (\cref{sec:mod_motor_system:anatomy}) as well as theoretical principles (\cref{sec:mod_motor_system:comp_modules_intro}). From a theoretical standpoint, we consider frameworks grounded in engineering principles (\cref{sec:mod_motor_system:engineering}), dimensionality-reduction approaches (\cref{sec:mod_motor_system:data_driven_output}), and dynamical systems perspectives (\cref{sec:mod_motor_system:data_driven_brain}). Despite addressing different facets of the motor control problem, operating at distinct levels of abstraction, and relying on different assumptions and data sources, these frameworks converge on a common conclusion: biological motor control is supported by a modular organizational structure.

\subsection{Anatomical modules}
\label{sec:mod_motor_system:anatomy}
Understanding the detailed mechanisms by which the motor system controls the body remains challenging, and identifying the specific contributions of individual neural structures is even more difficult. Nevertheless, certain neural structures possess well-defined neuroanatomical locations and functions that are broadly agreed upon in the classical literature \citep{kandel2021principles,grillner2023brain,lemon2008descending}.

\subsubsection{The spinal cord} The spinal cord occupies a central position in the motor system, as it contains alpha motor neurons. These excitatory neurons project directly to the skeletal muscle fibers, whose contractions generate joint motion. Thus, the spinal cord serves as the final common pathway for the generation of all body movements, with the exception of those involving neck and head muscles, which are triggered by alpha motoneurons in the brainstem. Additionally, the spinal cord contains several circuits that shape the spatial and temporal organization of muscle patterns underlying complex motor responses. Notable examples include the circuits that mediate reflexes and rhythmic movements, such as locomotion.

Spinal reflexes, which generate stereotyped motor responses to peripheral sensory inputs arising from receptors in muscles, skin, and joints, provide a classic example of spinal computation. These responses are mediated by circuits spanning a broad range of complexity. In the simplest case, monosynaptic pathways—such as those underlying the stretch reflex—allow proprioceptive afferents to project directly onto motor neurons. More complex polysynaptic circuits recruit networks of interneurons that coordinate multi-joint actions (e.g., withdrawal reflexes) or even bilateral, multi-limb responses (e.g., crossed extensor reflexes) \citep{kandel2021principles}. Importantly, reflex circuits are not hard-wired input–output mappings but can be strongly modulated by descending cortical and brainstem inputs, positioning the spinal cord as a major site of integration between peripheral sensory information and central motor commands \citep{scott2016functional}.

Consistent with this integrative role, substantial evidence indicates that the spinal circuits engaged during reflexive behaviors largely overlap with those recruited during voluntary movements; spinal interneuronal networks are not bypassed during voluntary control but are actively engaged \citep{nielsen2016human}. Moreover, the spinal cord contains circuits that, when activated, generate stereotyped spatiotemporal patterns of muscle activity corresponding to elementary submovements, even in the absence of proprioceptive input. These patterns—often referred to as motor primitives (\cref{sec:mod_motor_system:data_driven_output})—can be flexibly recruited, modulated, and combined by supraspinal centers to construct more complex, goal-directed behaviors.

\subsubsection{The brainstem} The brainstem is located directly above the spinal cord and connects it to the cerebrum. Like the spinal cord, it contains alpha motor neurons, but these innervate the neck and head muscles. Thus, the brainstem is the final pathway for the generation of all movements requiring the contraction of neck and head muscles. Similarly, it contains neural circuits responsible for various reflexes and rhythmic movements, such as breathing and chewing. 

Importantly, the brainstem also provides substantial input to the spinal cord. In some cases, these signals play a key role, finely coordinating the activation of multiple downstream spinal circuits with high temporal precision. This appears to be the case for behaviors such as reaching and grasping, where different brainstem nuclei appear to control the execution of specific movement phases \citep{esposito2014brainstem,ruder2019brainstem,ruder2021functional}. In other contexts, brainstem activity appears to exert a more modulatory influence. For example, well-characterized brainstem projections to spinal locomotor circuits are known to regulate parameters such as speed and gait, rather than fundamentally altering the underlying locomotor pattern itself \citep{leiras2022brainstem}.

A particularly specialized structure within the brainstem is the periaqueductal gray (PAG) \citep{bandler1994columnar}. The PAG contains spatially segregated circuits that orchestrate distinct survival-related motor behaviors, including flight, fight, freezing, foraging, and mating. These circuits can be directly engaged by inputs from the hypothalamus and amygdala, highlighting the PAG's role in facilitating the execution of instinctive behaviors critical for survival. In this respect, PAG-mediated actions differ from many other motor behaviors in that they are thought to be primarily triggered by internal states and innate drives rather than by voluntary, goal-directed control. At the same time, the PAG is embedded within broader recurrent networks. It receives input from cortical areas, potentially enabling voluntary or context-dependent engagement of survival behaviors, as well as from the basal ganglia, which may serve to suppress or gate PAG-driven actions that would be maladaptive in a given context. Importantly, although activation of PAG nuclei can initiate the expression of survival-related motor patterns, the full execution of these behaviors requires the recruitment of downstream brainstem and spinal cord circuits that house the premotor and motor neurons responsible for generating the precise spatiotemporal patterns of muscle activity characteristic of each behavioral phase \citep{grillner2023brain}.
 
\subsubsection{The primary motor cortex} \label{sec:primary_motor_ctx} The primary motor cortex (M1) is a cortical region located immediately anterior to the central sulcus and plays a central role in the execution of voluntary movements through its large number of projections to spinal and brainstem neurons. Other precentral regions --- the dorsal and ventral premotor cortices, the supplementary motor area, and the cingulate motor areas --- and parietal regions --- the primary somatosensory cortex, the posterior parietal cortex, and the parietal operculum ---  also project directly to spinal and brainstem interneurons \citep{lemon2008descending}. However, M1 is the only region containing a special class of neurons --- named corticomotoneuronal (CM) cells --- that project monosynaptically to spinal alpha motor neurons. CM cells are thus able to directly influence the contractions of their target muscles without having to rely on interneurons and reflex circuits that the interneurons are part of. Interestingly, CM cells are only found in higher primates (e.g., humans, macaques, and capuchin monkeys), are directly linked to dexterity, and cluster in the caudal region of M1. This region, sometimes referred to as new M1, can be considered as a special module within M1 that evolved to provide finer control over hand and arm muscles: bypassing spinal reflex circuits, it enables skillful movements and thus unprecedented dexterity \citep{rathelot2009subdivisions}.

Given that cortico-motoneuronal (CM) cells are found almost exclusively in M1, that they predominantly target hand and finger muscles, and that they are present only in higher primates—with their number scaling with manual dexterity—there is broad consensus that M1 plays a central role in the control of fine hand movements, including fractionated finger actions \citep{lemon2008descending,courtine2007can}. Moreover, the strong anatomical and functional coupling between M1 and sensory regions of the parietal cortex underscores its importance for actions that depend on detailed sensory information, such as object manipulation, tool use, and visually guided movements. In contrast, the contribution of M1 to the control of other movement types is less well understood and remains an active area of investigation.

Indeed, although all premotor areas project to M1, and M1 is a major source of descending projections to the brainstem and spinal cord, it is not the sole cortical structure capable of influencing voluntary behavior. Multiple other cortical regions—including premotor areas and even primary somatosensory cortex (S1)—also project directly to downstream brainstem and spinal circuits. For example, according to some estimates, only 1/3 of the cortical input to the spinal cord comes from M1 \citep{strick2021cortical}. Thus, contrary to a common but oversimplified view, M1 is not the only cortical node through which voluntary actions can be expressed. Consistent with this perspective, animals with surgical lesions of the corticospinal tract (but with an intact corticobulbar tract, which carries cortical axons to the brainstem) retain a wide repertoire of intact behaviors and appear to be primarily impaired in the fine control of distal hand movements \citep{lawrence1968functional}. This observation aligns with evidence that many brainstem \citep{esposito2014brainstem} and spinal cord circuits \citep{alstermark2012circuits} possess the necessary machinery to generate coordinated patterns of muscle activity underlying ethologically relevant behaviors such as reaching and grasping. Remarkably, even when decorticated, animals, such as cats, retain most of their behavioral repertoire \citep{bjursten1976behavioural}; similarly, rodents with motor cortical lesions are often perfectly capable of executing pre-lesion movements, even if recently learned, despite exhibiting marked deficits in learning new motor skills \citep{kawai2015motor}.

Nevertheless, M1 has occupied a central position in motor control research since the earliest electrical stimulation studies of the 19th century \citep{fritsch1870uber,ferrier1873experimental}, a prominence that has persisted through decades of work aimed at characterizing the response properties of M1 neurons. A long-standing debate has focused on whether M1 activity primarily reflects low-level motor variables, such as muscle force \citep{evarts1968relation}, or higher-level, more abstract parameters, such as movement direction \citep{georgopoulos1982relations}. This debate began to settle when it was shown that M1 contains distinct classes of neurons whose activity preferentially reflects either muscle-related or kinematic variables \citep{kakei1999muscle}. More recently, M1 has remained a focal point of investigation, but the emphasis has shifted away from identifying the specific variables encoded by individual neurons toward a dynamical systems perspective (\cref{sec:mod_motor_system:data_driven_brain}). In this view, population-level activity in M1 is analyzed to uncover the computations and transformations that support movement generation, rather than to assign fixed representational meanings to average firing patterns of individual neurons \citep{shenoy2013cortical,sabatini2024reach}.

\subsubsection{The premotor cortex} \label{sec:premotor_ctx} The premotor cortex, which lies rostrally to M1, immediately adjacent to it, is traditionally believed to play a central role in the planning and selection of voluntary movements. Like M1, it lacks the internal granular layer IV, which typically provides the cortex with sensory inputs from the thalamus. Unlike M1, the premotor cortex lacks the large pyramidal cells in layer V (Betz cells). The three most well-characterized subregions of the premotor cortex are the dorsal premotor cortex (PMd), the ventral premotor cortex (PMv), and the supplementary motor area (SMA) \citep{dum2002motor}. These regions contain reciprocal connections with each other and with M1 that are somatotopically organized. 

While both SMA and M1 receive somatotopically organized sensory information from the primary sensorimotor cortex and immediately adjacent parietal regions, PMd and PMv have recurrent connections with higher-level regions of the parietal cortex. In addition to inputs from M1 and parietal cortex, each premotor region also receives a unique pattern of input from the prefrontal cortex and is reciprocally connected with different basal ganglia and cerebellum circuits \citep{dum2002motor}. Such distinctive connectivity patterns are paralleled by distinctive functions ascribed to the premotor areas. For example, while PMd and PMv are often involved in actions generated in response to sensory inputs, SMA is more often involved in self-initiated actions (similarly to other important premotor regions, including the presupplementary and cingulate motor areas). Additionally, while PMd and SMA appear to be predominantly involved in planning arm movements, PMv appears to be mostly involved in planning hand movements. 

\subsubsection{The posterior parietal cortex}
\label{sec:ppc} The posterior parietal cortex (PPC) is located between the primary somatosensory cortex and the occipital lobe and is believed to integrate somatosensory, visual, vestibular, and auditory inputs to build a modality-independent representation of the state of the body and the world to guide the generation of appropriate motor commands. PPC projects to both cortical areas, including the premotor and prefrontal cortices, and to subcortical structures such as the spinal cord \cite{rathelot2017posterior}. Given the multimodal and heterogeneous nature of its activation dynamics, PPC is believed to subserve multiple functional roles. Within PPC, extensively studied regions are the lateral intraparietal area (LIP),
the ventral intraparietal area (VIP),
the medial intraparietal area (MIP), Broadmann's area 5 (PE), and the PE intraparietal area (PEip). 

These regions receive somatosensory inputs from the primary somatosensory cortex conveying information about body posture and movement, as well as visual inputs from the dorsal visual stream encoding the physical attributes of objects in the visual field, including their location, shape, and size.

Neurons in these areas exhibit consistent activation dynamics during arm and hand movements, which are thought to reflect representations of visual stimuli in distinct reference frames. On the basis of these response properties, LIP is proposed to encode retina-centered representations, VIP head-centered representations, whereas PE and PEip primarily encode hand-centered representations. Interestingly, MIP appears to encode hybrid, retina-centered representations of reach direction, which depend on the location of both the visual stimuli and the hand.

Another important region within the posterior parietal cortex is the anterior intraparietal area (AIP). This region is particularly sensitive to the shape, orientation, and size of objects that afford interaction, suggesting a key role in grasping and object manipulation.

The response properties of these posterior parietal regions are consistent with their known patterns of connectivity with premotor areas. For example, AIP, which is preferentially involved in hand-related functions, is more strongly connected with ventral premotor cortex (PMv), whereas PE and MIP, which are preferentially involved in arm-related functions, show stronger interactions with dorsal premotor cortex (PMd) and supplementary motor area (SMA).

\subsubsection{The cerebellum}
\label{sec:anatomy:cerebellum}
The cerebellum is a large subcortical brain structure located at the posterior aspect of the brain, beneath the occipital lobes, within the posterior cranial fossa. Notably, it is the brain structure with the highest neuronal density: although it occupies only about 10\% of total brain volume, it contains approximately 80\% of all neurons in the brain \citep{van2018development}. It is reciprocally connected with several cortical regions and with the basal ganglia and plays a central role in motor coordination, motor learning, and motor timing. In addition to its contribution to the motor system, it is also clearly involved in a range of non-motor functions, including executive control, language, and emotional processing. Anatomically, the cerebellum consists of a highly folded cerebellar cortex—the principal site of cerebellar computation—an underlying layer of white matter, and three pairs of deep cerebellar nuclei, which constitute its main output structures: the fastigial, interposed, and dentate nuclei. 

Functionally, the cerebellum comprises three principal domains: the vestibulocerebellum, spinocerebellum, and cerebrocerebellum, each providing a different contribution to distinct classes of movements. 
The vestibulocerebellum domain integrates visual and vestibular sensory signals and exerts its influence primarily through projections to the vestibular nuclei of the brainstem, where it contributes to the control of balance, eye movements, and vestibular reflexes. 
The spinocerebellum domain receives strong proprioceptive inputs from the spinal cord, as well as vestibular, visual, and auditory information; it projects to cortical and brainstem regions and contributes to the control of posture, locomotion, and eye movements.
The cerebrocerebellum domain constitutes the largest subdivision of the cerebellum and is reciprocally connected with multiple cortical regions, including primary motor, premotor, and posterior parietal cortices. Its output to the cortex is conveyed via the thalamus, while its cortical input is relayed through the brainstem. In addition to projecting to cortical regions, the cebrocerebellum also sends projections to the spinal cord via the brainstem. Although this domain contributes to a broad range of functions, it plays a central role in motor planning and coordination and is additionally implicated in non-motor processes, such as language and visuospatial cognition.

In motor control, the cerebellum has long been implicated in ensuring coordinated movement. One influential account posits that the cerebellum contributes to coordination by estimating and compensating for interaction torques arising from the mechanical couplings between adjacent limb segments. For example, when flexing the elbow, compensatory activation of shoulder muscles is required to stabilize the arm and prevent unwanted motion at adjacent joints. Healthy individuals perform this compensation efficiently, enabling smooth multi-joint movements. In contrast, patients with cerebellar dysfunction often fail to anticipate interaction torques, leading to unintended, task-irrelevant joint movements \citep{bastian1996cerebellar,maeda2017compensating,oh2025cerebellar}. As a consequence, they exhibit pronounced deficits in multi-joint coordination and frequently decompose complex movements into sequences of simpler, single-joint actions as a compensatory mechanism. 

How the cerebellum generates motor commands that compensate for interaction torques remains an active area of investigation. A widely held view is that it does so by leveraging {\it internal models} of the body and the external environment \citep{wolpert1998internal}. These models enable the motor system not only to generate compensatory motor commands that stabilize joints against mechanical coupling, but also to support more general computations. In particular, internal models are thought to estimate how motor commands will alter the state of the body ({\it forward models}) \citep{miall2007disruption,izawa2012cerebellar} and to compute the motor commands most appropriate for achieving a desired change in body state ({\it inverse models}) \citep{flanagan2003prediction,honda2018tandem}.

The cerebellum is also known to play a fundamental role in error-based motor learning, which involves adapting motor commands to compensate for changes in the mapping between motor commands and their sensory consequences, which perturb natural behavior. Such changes may arise naturally, for example, through alterations in muscle properties, or may be experimentally imposed via mechanical perturbations. In neurologically intact individuals—but typically not in cerebellar patients \citep{martin1996throwing,therrien2015cerebellar}—these perturbations are gradually accounted for by adaptive changes in motor commands that reestablish baseline performance. 

The learning signal driving motor adaptation is widely believed to be the so-called {\it sensory prediction error}, defined as the mismatch between predicted and actual sensory feedback. This signal is thought to drive the recalibration of both forward and inverse internal models \citep{mazzoni2006implicit,tseng2007sensory,shadmehr2010error}. The circuit mechanisms underlying error-based learning in the cerebellum remain incompletely understood. However, substantial evidence suggests that a large fraction of this learning occurs within the cerebellar cortex, the principal computational substrate of the cerebellum. The primary processing units of the cerebellar cortex are Purkinje cells and granule cells, while the main cerebellar inputs are sensory signals provided by climbing fibers—originating from neurons in the inferior olive in the brainstem—and sensory-motor signals provided by mossy fibers, arising from cell bodies in the spinal cord and brainstem. Mossy fibers are thought to convey information related to recently issued motor commands. This information is relayed to granule cells and subsequently transmitted to Purkinje cells through the granule cell axons, which form long parallel fibers spanning the cerebellar cortex. Activity in this pathway drives simple spike firing in Purkinje cells. In contrast, climbing fiber activity is widely believed to reflect sensory prediction errors. When such an error is detected, climbing fibers fire and elicit complex spikes in Purkinje cells, characterized by a large initial depolarization followed by a burst of smaller spikelets in the soma and dendrites. The occurrence of a complex spike induces synaptic depression at parallel fiber–Purkinje cell synapses that were active shortly beforehand. These parallel fibers are therefore inferred to have contributed to the erroneous motor command that generated the prediction error. Therefore, climbing fibers gate error-specific synaptic changes, biasing plasticity toward weakening synapses that contributed to erroneous motor outputs, while complementary potentiation and homeostatic mechanisms ensure more appropriate Purkinje cell simple spike output patterns \citep{herzfeld2018encoding,herzfeld2020principles}.

\subsubsection{The basal ganglia} \label{sec:basal_ganglia}
The basal ganglia are a group of interconnected subcortical nuclei located deep within the brain, surrounding the uppermost portion of the brainstem. In addition to interacting with the brainstem, they influence motor and non-motor areas of the cerebral cortex, as well as limbic structures such as the hippocampus and the amygdala. Most of the connections are recurrent and are mediated by thalamic relay nuclei. Functionally, the basal ganglia contribute critically to motor control—particularly the initiation and selection of actions—and to reward-based learning, while also modulating motivational, cognitive, and affective processes.
The three main input structures of the basal ganglia (BG) are the striatum, the subthalamic nucleus (STN) and the substantia nigra pars compacta (SNc), while the two main output structures are the internal globus pallidus (GPi) and substantia nigra pars reticulata (SNr). The external globus pallidus (GPe) mostly contains connections with other BG structures and is thus considered an intrinsic nucleus.
GPi and SNr contain tonically active, GABAergic, inhibitory neurons that project to brainstem motor nuclei and thalamus. Therefore, at rest, BG keep their target brainstem and thalamic regions inhibited. 

The striatum contains two classes of GABAergic, inhibitory spiny striatal projection neurons (SPNs): D1-SPNs, expressing D1 receptors, and D2-SPNs expressing D2 receptors. The main functional difference between these two classes of neurons \footnote{It is known that while D1 receptors are excited by the neurotransmitter dopamine, D2 receptors are inhibited by it; however, the functional relevance of these receptors is currently debated.} is that while D1-SPNs project to the BG output nuclei directly, and thus have a net excitatory effect on the target BG regions, D2-SPNs project to the BG output nuclei indirectly, via the inhibitory neurons of GPe and thus have a net inhibitory effect on the downstream areas targeted by BG projections. The existence of these two complementary pathways --- the {\it direct} one mediated by D1-SPNs, and the {\it indirect} one mediated by D2-SPNs --- has historically been central to influential models of BG function, in which the BG circuits are thought to translate high-level movement specifications into the precise pattern of disinhibition and inhibition across subcortical motor centers required to execute the desired action. Upon receiving a cortical input coding for the desired movement, the direct pathway disinhibits the motor centers required for that movement, while the indirect pathway strongly suppresses the motor centers which would elicit inappropriate or interfering submovements \citep{grillner2016basal}.

The prevailing view is that the basal ganglia do not merely implement a fixed mapping between desired actions and the motor centers required to execute them. Rather, they are thought to play a central role in action selection. Evidence suggests that the basal ganglia participate in recurrent loops with their principal input structures, particularly cortex and thalamus, projecting via relay nuclei back to the same regions from which they receive input. Through these closed-loop circuits, the basal ganglia can shape cortical and thalamic activity by selectively inhibiting or disinhibiting specific input channels, effectively providing a filtered version of incoming signals. According to this framework, the basal ganglia receive inputs representing multiple competing behavioral options that are plausible given the current sensory context and, through internal circuitry shaped by past reward history, select the most appropriate option. This selected signal is then relayed back to upstream regions (i.e., motor cortex and brainstem), which can propagate the chosen action to downstream motor systems for execution.

Within this framework, dopamine (DA) is thought to play a fundamental role in modulating BG excitability, information gating—by permitting selected signals to be transmitted downstream—and learning. Dopaminergic neurons originate primarily from two midbrain nuclei: the substantia nigra pars compacta (SNc), which projects predominantly to the dorsolateral striatum, and the ventral tegmental area (VTA), which innervates more medial striatal regions. Under baseline conditions, both populations exhibit tonic DA release, maintaining striatal circuits in a sufficiently excitable state. Superimposed on this tonic activity, phasic DA release from the SNc has been observed immediately prior to movement initiation and appears to be necessary for the selected action to be transmitted to downstream motor circuits and executed. Moreover, the magnitude of this phasic dopaminergic signal correlates with movement vigor, with larger DA transients associated with faster or more forceful actions. 
In contrast, dopaminergic neurons in the ventral tegmental area (VTA) exhibit robust phasic firing in response to unexpected rewards or unexpected reward-predicting cues. This VTA-derived dopaminergic signal is widely thought to support learning of new, better, behavioral responses to a given sensory input that are more likely to be rewarded. Such VTA-mediated plasticity is believed to occur primarily at corticostriatal synapses in the medial striatum, as well as within cortical circuits themselves \citep{arber2022networking}.

More broadly, the selection mechanisms implemented by the basal ganglia circuits are thought to extend beyond motor control. Rather than being restricted to action selection in cortical motor areas, similar basal ganglia–mediated computations are believed to operate across multiple cortical and subcortical domains, supporting the selection of appropriate cognitive representations, motivational goals, or limbic states depending on behavioral context \citep{matsumoto2013distinct}.

\subsubsection{An overall view}
\label{sec:anatomy:overall_view}
In the sections above, we have outlined—largely in isolation—the structure and function of the motor system components most relevant for sensorimotor control, progressing from structures that are anatomically and functionally closer to motor output (such as the spinal cord) to those that are more distant and strongly shaped by high-level goals and sensory representations of the environment (such as the posterior parietal cortex). This overview is necessarily cursory and highly schematic: each of these structures comprises multiple subregions, nuclei, or distributed circuits, with context-dependent activity and diverse functional roles. Furthermore, although these components are spatially segregated and often discussed in terms of distinct functions, they interact tightly through recurrent and overlapping circuits. These interactions give rise to distributed computations that are essential for generating flexible, accurate, and smooth movements.

For example, recent work has further clarified the extent to which cortex, basal ganglia, and cerebellum are tightly interconnected. It has long been established that both the basal ganglia and the cerebellum form closed-loop circuits with motor and non-motor regions of the cortex. For many years, however, it was assumed that information from these two subcortical systems could be integrated only at the cortical level (via cortico-cortical loops) after being relayed through distinct thalamic nuclei. More recent anatomical and functional evidence has overturned this view, revealing disynaptic subcortical connections between the basal ganglia and the cerebellum that allow them to influence one another independently of the cortex \citep{bostan2018basal}. Specifically, neurons in the cerebellar dentate nucleus project to the striatum—preferentially targeting the indirect pathway—via the thalamic central lateral nucleus. Conversely, the basal ganglia subthalamic nucleus ---a central node in the indirect pathway--- projects to the cerebellar cortex through the brainstem pontine nuclei. 

The computational roles of these interconnections remain largely unresolved. Several theoretical frameworks have proposed possible interpretations, but direct empirical evidence is still limited. One class of hypotheses \citep{doya1999computations,caligiore2017consensus} posits that signals sent from cortex to cerebellum convey simulated motor commands whose sensory consequences must be predicted, consistent with the cerebellum’s established role in forward modeling, or motor commands reflecting the current motor output, which the cerebellum relies on for error-based learning. In contrast, signals sent from the cortex to the basal ganglia may primarily be used to select among or evaluate the expected reward from candidate motor commands. The functional significance of cerebellar signals reaching the basal ganglia is similarly open to interpretation. Such signals may represent simulated motor commands whose predicted consequences must be evaluated, or they may convey information about the temporal structure of planned movements that the basal ganglia can use to coordinate action initiation and termination. Likewise, the role of signals sent from the basal ganglia to the cerebellum is particularly difficult to infer. One possibility is that they convey a suppressive or terminating signal, instructing the cerebellum to disengage ongoing motor corrections. Another possibility is that they signal the absence of an ongoing action, thereby allowing the cerebellum to operate in an offline or predictive mode, such as simulating the consequences of potential future actions.

How, then, do the different computational nodes of the motor system interact to produce context-dependent and accurate behavior? A first step is to consider what computational and anatomical resources are available to the motor system. It is well established that brainstem \citep{esposito2014brainstem} and spinal cord \citep{alstermark2012circuits} circuits are capable of generating a rich repertoire of ethologically relevant behaviors \citep{grillner2023brain}. In primates, this baseline architecture is complemented by highly developed corticobulbar and corticospinal projections \citep{lemon2008descending,rathelot2009subdivisions}, through which cortical areas can directly influence subcortical motor circuits or modules and even the activity of individual alpha motoneurons. Together, these pathways provide a substrate for both executing existing behaviors --- by recruiting existing modules --- and for fine-tuning them --- by shaping the activation dynamics of existing modules with corticospinal projections.

Recruitment of existing movement-coding subcortical motor modules is likely mediated by corticobulbar projections from the dorsal premotor \citep{graziano2002complex,desmurget2009movement} or posterior parietal \citep{andersen2002intentional,stepniewska2005microstimulation} regions, which have been shown to contain representations of complete, ethologically relevant behaviors. In contrast, fine-grained shaping of activity patterns within these modules—particularly at the level of individual muscles or digits—likely requires more direct cortical access and therefore depends more strongly on corticospinal projections from the primary motor cortex \citep{griffin2015corticomotoneuronal}.

Within this framework, motor learning can be understood as the process by which the nervous system modifies and extends its existing repertoire of movement-generating modules. Learning may occur through the formation of new combinations of preexisting modules, the fine-tuning of existing modules, or, in some cases, the creation of entirely new ones \citep{logiaco2021thalamic}. The underlying mechanisms remain incompletely understood, as multiple brain structures contribute to these processes \citep{krakauer2019motor}. Current evidence indicates that motor learning depends on plasticity distributed across cortical and subcortical circuits, with the motor cortex \citep{kawai2015motor}, basal ganglia \citep{doyon2009contributions}, and cerebellum \citep{ito2000mechanisms} each making distinct but complementary contributions to the acquisition and long-term storage of motor skills \citep{strick2021cortical}.

Given this architecture, how are movement plans actually generated? This question remains open. An influential proposal, often referred to as the affordance competition hypothesis \citep{cisek2007cortical,cisek2010neural}, suggests that the motor system continuously represents not only the salient objects present in the environment, but also the actions that those objects afford. These potential actions are thought to be represented within frontoparietal circuits as candidate motor plans. According to this framework, multiple action representations coexist and compete with one another. Their relative expression is dynamically biased through recurrent interactions with a variety of cortical and subcortical systems until one representation becomes dominant. For example, interactions with prefrontal cortex may bias candidate actions according to abstract, high-level goals and contextual demands \citep{miller2000prefontral}, interactions with the cerebellum may bias them based on predicted sensory consequences, and interactions with the basal ganglia may bias them according to expected reward. Through these complementary biasing processes, competition among candidate actions is progressively resolved. Once a clear winner emerges, downstream subcortical motor centers associated with the selected action are engaged, and the action is executed.

\begin{tcolorbox}[
  enhanced,
  breakable,
  colback=orange!10,
  colframe=orange!60!black,
  title=Adaptation and learning in the motor system,
  before skip=0.6\baselineskip,
  after skip=0.6\baselineskip
]
\label{box:adaptations_learning}
A popular strategy for understanding how the sensorimotor system handles naturally occurring changes in body and world dynamics is to simulate these changes using artificial perturbations in experimental settings \citep{von1925treatise,lackner1994rapid,shadmehr1994adaptive,li2001neuronal}. Such perturbations disrupt ongoing motor behavior by altering the relationship between issued motor commands and their sensory consequences. The sensorimotor system then senses this alteration by detecting the discrepancy between expected and experienced sensory consequences resulting from the perturbation \citep{shadmehr2010error}, and tries to correct motor commands to reestablish dexterous behavior. Healthy subjects typically adapt quickly to reestablish an acceptable level of performance even in this perturbed environment. In line with previous studies \citep{li2001neuronal,smith2006interacting,joiner2008long}, we refer to this phenomenon as \emph{short-term learning} or \emph{adaptation}. When the perturbation is removed, subjects gradually deadapt their motor commands and return to baseline behavior. Interestingly, if they are subsequently exposed to the same perturbation again, they will generally be able to adapt their motor commands faster than they did during the first exposure. This phenomenon, known as \emph{savings} \citep{ebbinghaus2013memory}, has been robustly observed in several motor behaviors \citep{brashers1996consolidation,krakauer2005adaptation,roemmich2015two} and shows strong resilience to the passage of time as it persists even one year after the first exposure \citep{landi2011one}. Savings is therefore the evidence that the exposure to a single transient motor perturbation is strong enough to induce permanent changes in the sensorimotor system and is thus a signature of \emph{long-term learning} or simply \emph{learning}. The standard method for measuring learning is to compare adaptation and readaptation dynamics. An alternative method, often preferred for convenience, is to quantify how quickly subjects return to baseline behavior after the first perturbation is removed. During this phase, subjects typically continue to apply the altered motor commands even after the perturbation has been removed. This will lead to the so-called aftereffect errors, and their magnitude is assumed to be proportional to the amount of learning \citep{shadmehr1994adaptive}. 
\\

Are adaptation and learning two sides of the same coin? Are they mediated by a common neural system, or do they reflect distinct computational processes? Moreover, can learning occur in the absence of adaptation? While definitive answers to these questions remain elusive, several lines of evidence offer important clues. A large body of evidence suggests that, regardless of the details of the experimental settings, adaptation and learning reflect the operations of qualitatively distinct systems. For example, the magnitude of the errors elicited by a perturbation seems to mostly modulate the amount of learning without significantly impacting the adaptation dynamics \citep{roemmich2015two,morehead2015savings}. Likewise, providing a reward during adaptation significantly affects learning \citep{izawa2012cerebellar}, but only slightly affects the adaptation rate \citep{galea2015dissociable}. Conversely, providing augmented error feedback \citep{long2016blocking,roemmich2016seeing} or explicit instructions \citep{malone2010thinking} during adaptation significantly affects the rate of adaptation but does not affect learning. Thus, adaptation dynamics are often at odds with learning dynamics, which are more strongly influenced by error size and reward signals. Further proof of this dichotomy between adaptation and learning is provided by studies that modulate the excitability of key motor regions during the execution of sensorimotor learning tasks \citep{galea2011dissociating,richardson2006disruption,hadipour2007impairment}. Specifically, such studies consistently report that increasing the excitability of the primary motor cortex through anodal transcranial direct current stimulation (tDCS) or decreasing it through cathodal tDCS or repetitive transcranial magnetic stimulation (rTMS) affects learning but not adaptation. However, similar cerebellar stimulations affect both adaptation and learning \citep{galea2011dissociating,koch2020improving}. Thus, while cerebellar processes appear to contribute to both adaptation and learning, primary motor cortex processes appear to contribute more strongly to learning. The precise relationship between adaptation and learning remains unresolved. Nevertheless, evidence from clinical studies \citep{weiner1983adaptation,martin1996throwing,marinelli2009learning,leow2012impaired,leow2013different} is consistent with the hypothesis that adaptation is \emph{necessary} but not \emph{sufficient} to induce learning. Cerebellar patients, for example, are typically unable to adapt their motor commands and show little to no evidence of learning \citep{weiner1983adaptation,martin1996throwing}. In contrast, patients with Parkinson’s disease generally retain the ability to adapt but exhibit pronounced deficits in learning \citep{marinelli2009learning,leow2012impaired,leow2013different}. Together, these dissociations suggest the involvement of at least two distinct systems. Notably, this distinction parallels the well-established separation between performance and learning in the cognitive psychology literature \citep{cowan2008differences,kantak2012learning,soderstrom2015learning}.
\\

Taken together, these findings suggest that adaptation and learning in the human sensorimotor system are mediated by qualitatively distinct yet interacting systems. Nevertheless, the sensorimotor system is often described in terms of a single underlying mechanism—typically linear and time-invariant—that is sometimes augmented with multiple, qualitatively similar processes \citep{thoroughman2000learning,smith2006interacting,inoue2015three}. These processes are generally distinguished by their time constants and are assumed to work in concert to improve performance. While such models capture key aspects of short-term adaptation, they fail to account for longer-term learning phenomena, such as savings \citep{zarahn2008explaining}, and cannot readily explain the differential effects of error magnitude and reward on adaptation and learning. Collectively, the studies reviewed here instead suggest that a comprehensive model of sensorimotor adaptation and learning should explicitly represent these two components as disentangled and qualitatively different. In such a framework, adaptation may more closely reflect fast, error-driven processes associated with cerebellar circuitry, whereas learning may arise from slower processes mediated by basal ganglia and motor cortical circuits that stabilize and consolidate corrective signals through longer-term plasticity mechanisms. Finally, we speculate that while cerebellar-dependent corrective processes are always engaged to quickly re-establish an acceptable level of performance, basal ganglia- and motor cortex-dependent learning processes are engaged only when sensory prediction errors are sufficiently large.

\end{tcolorbox}

\subsection{Computational modules: a recurrent motif in motor control theories}
\label{sec:mod_motor_system:comp_modules_intro}
As discussed in the preceding sections (\cref{sec:mod_motor_system:anatomy}), from a neuroanatomical perspective the motor system is composed of multiple distinct modules, each characterized by a specific anatomical location, connectivity profile, and activation dynamics that covary with particular movement features across different movement phases. This modular organization has motivated attempts to assign distinct functions to individual components of the motor system. However, substantial overlap between modules remains evident, and explaining how these components cooperate to generate coordinated behavior has proven challenging (\cref{sec:anatomy:overall_view}). In other words, it is difficult to arrive at a unifying account of motor control through a purely bottom-up approach that builds on the isolated study of the anatomy, connectivity, and functional properties of individual neural structures. Although such approaches are invaluable, they often fall short of explaining how distributed neural components collectively give rise to coherent and adaptive behavior.

By contrast, computational theories of motor control typically adopt a top-down perspective. These approaches begin with normative or computational principles specifying how an effective motor system should operate to support accurate and flexible control under computational constraints, and then assess whether the resulting predictions are consistent with empirical data. Predictions derived from these theories are subsequently evaluated, refined, or rejected through carefully designed behavioral and neurophysiological experiments.

In the following sections, we focus on three influential computational theories of motor control. Despite important differences, all three can be viewed as fundamentally modular frameworks that posit distinct computational components. For example, in optimal feedback control theory (\cref{sec:mod_motor_system:engineering}), modules are responsible for system identification, state estimation, cost computation, and state-dependent control. In motor-primitive–based theories (\cref{sec:mod_motor_system:data_driven_output}), modules encode reusable spatial and temporal coordination patterns that mitigate musculoskeletal redundancy and reduce the dimensionality of muscle control signals. Finally, in dynamical systems theories of motor control (\cref{sec:mod_motor_system:data_driven_brain}), modules enable the functional decoupling of motor preparation, timing, and execution, allowing each process to be optimized independently while supporting robustness to noise and contextual flexibility.

\subsection{Functional modules for state-dependent motor control}
\label{sec:mod_motor_system:engineering}
In the following sections, we examine optimal feedback control (OFC) theory \citep{todorov2002optimal} and its core computational modules, which recur across many influential theories of motor control (e.g., \cite{wolpert1995internal,harris1998signal,wolpert1998multiple}). In two seminal articles, these modules were first integrated into a unifying computational framework by OFC theory \citep{todorov2002optimal} and were subsequently, albeit tentatively, associated with specific brain areas \citep{shadmehr2008computational}. In short, OFC theory provides a mechanism by which the motor system can exploit delayed state feedback to generate motor commands flexibly, in a state-dependent manner, without specifying a desired trajectory a priori. This mechanism relies on a computationally demanding learning phase to acquire an optimal policy, but it reduces online computational demands during movement execution, as only behaviorally relevant deviations from the task goal need to be corrected.

\subsubsection{The system identification module}
This module learns a {\it forward model} of the dynamics of the body and environment. Specifically, it estimates how motor commands $\mathbf{u}(t) \in \mathbb{R}^m$ affect the state of the system $\mathbf{x}(t) \in \mathbb{R}^n$ and how that state maps onto sensory feedback $\mathbf{y}(t) \in \mathbb{R}^f$:
\begin{align}
\hat{\mathbf{x}}{(t+1|t)} &= \hat{A}\,\hat{\mathbf{x}}{(t)} + \hat{B}\,\mathbf{u}{(t)} \\
\hat{\mathbf{y}}{(t)} &= \hat{H}\,\hat{\mathbf{x}}{(t)}
\end{align}
where $\hat{A}$, $\hat{B}$, and $\hat{H}$ are the estimated transition, input and output matrices, respectively.
By predicting the sensory consequences of motor commands prior to the arrival of delayed sensory feedback, the forward model supports rapid detection and correction of motor errors. Such predictive processes are particularly critical for fast movements, such as saccades, in which corrective adjustments must be initiated before sensory feedback can be processed. More generally, this sort of prediction is required whenever motor commands to a given muscle group must account for the anticipated sensory consequences of commands issued to other muscle groups—a ubiquitous requirement given the mechanical coupling inherent to the musculoskeletal system. 

As we have discussed in the previous sections, there is substantial evidence that the cerebellum is strongly implicated in storing (c.f., \cref{sec:anatomy:cerebellum}) and updating forward models (c.f., \cref{box:adaptations_learning}). In particular, patients with cerebellar damage exhibit marked impairments in predicting the sensory consequences of their motor commands and in adapting internal models to changing dynamics \citep{therrien2015cerebellar,wolpert1998internal}. Therefore, cerebellar circuits are well poised to serve as a neural substrate for system identification and forward modeling.

\subsubsection{The state estimation module}
This module combines prior predictions with incoming sensory information to form a reliable estimate of the current state of the body and environment. Formally, it computes a posterior estimate $\hat{\mathbf{x}}{(t+1|t+1)}$ by integrating the predicted state $\hat{\mathbf{x}}{(t+1|t)}$ with the discrepancy between observed and predicted sensory feedback:
\begin{align}
\hat{\mathbf{x}}{(t+1|t+1)} =
\hat{\mathbf{x}}{(t+1|t)} +
K{(t+1)}\bigl(\mathbf{y}{(t+1)} - \hat{\mathbf{y}}{(t+1)}\bigr) .
\end{align}
where $K(t+1)$ is a mixing gain matrix that weights how much we should trust the sensory feedback to correct our prior estimate. It was shown that this state estimation process in human subjects can be well approximated by Kalman filtering \citep{kalman1961new}, where $K(t+1)$ is determined so that $\hat{\mathbf{x}}{(t+1|t+1)}$ is the maximum likelihood estimate of the state, assuming Gaussian and uncorrelated state and observation noise \citep{wolpert1995internal,kording2004bayesian}. This integration yields a state estimate that is more accurate and robust than either predictions or sensory signals alone, flexibly upweighting the contributions of signals that appear more reliable. 

As we have previously discussed (c.f., \cref{sec:ppc}), the posterior parietal cortex is believed to integrate multiple sensory inputs to create useful representations of the body's and the world's state, guiding the generation of appropriate motor commands. In general, permanent or transient damage to the posterior parietal cortex impairs the rapid integration of new sensory information. For example, when individuals with parietal damage are instructed to reach toward a target that unexpectedly changes location during movement execution, they tend to continue reaching toward the initial target, failing to incorporate the updated sensory information available mid-reach \citep{desmurget1999role,grea2002lesion}. Specifically, inferior parietal lobe (IPL) regions appear to be mostly involved with the creation of representations that depend on visual inputs, while superior parietal lobe (SPL) regions appear to be mostly dependent on proprioceptive inputs \citep{rushworth1997parietal}.

\subsubsection{The cost module}
This module estimates the instantaneous cost associated with being in a particular state $J_{{\mathbf{x}}}(t,\hat{\mathbf{x}}(t))$, and the instantaneous cost associated with the generation of a particular motor command $J_{\mathbf{u}}(t,\mathbf{u}(t))$. For example, when a task rewards tracking a reference trajectory $\mathbf{r}(t)$ and penalizes motor effort, instantaneous state and movement costs may be expressed as
\begin{align}
J_{{\mathbf{x}}}(t,\hat{\mathbf{x}}(t)) 
&= \bigl(\hat{\mathbf{x}}(t)-\hat{\mathbf{r}}(t)\bigr)^{\mathsf T}
\hat{S}(t)\,
\bigl(\hat{\mathbf{x}}(t)-\hat{\mathbf{r}}(t)\bigr) \\
J_{\mathbf{u}}(t,\mathbf{u}(t)) 
&= \mathbf{u}(t)^{\mathsf T} \hat{M}(t)\,\mathbf{u}(t)
\end{align}
where $\hat{\mathbf{r}}(t)$ is the estimated reference trajectory, while $\hat{{S}}$ and $\hat{{M}}$ are the estimated state and movement cost matrices, respectively. These cost functions formalize task goals and energetic constraints, defining the objective that guides motor behavior.

Empirical evidence indicates that larger motor commands are associated with increased motor noise, leading to reduced movement accuracy, thereby favoring the use of lower-amplitude commands \citep{harris1998signal}. In addition, larger motor commands incur greater metabolic costs. Consistent with these constraints, humans tend to prefer less effortful movements, and models that minimize metabolic cost yield reaching trajectories that closely resemble those observed in natural behavior \citep{alexanderRM97}. Notably, perceived effort has been shown to scale approximately quadratically with generated force \citep{morel2017makes}. For these reasons, and for mathematical convenience, human sensorimotor decision-making is often modeled using quadratic cost functions—partly; interestingly, however, empirical evidence suggests that subjective loss functions may deviate from strict quadratic forms, particularly for large values of their domain \citep{kording2004loss}. 

Damage to the basal ganglia, as commonly observed in Parkinson’s disease, typically results in slow, low-amplitude movements (bradykinesia). These deficits are often attributed to impaired disinhibition of downstream motor centers arising from reduced tonic and phasic dopamine release from the substantia nigra pars compacta (SNc). Diminished dopaminergic signaling limits baseline excitability within the striatum and compromises its ability to effectively propagate pre-movement thalamocortical signals to basal ganglia output nuclei, thereby impairing movement initiation and reducing movement vigor (\cref{sec:basal_ganglia}). However, slow, low-amplitude movements are also compatible with increased motor costs associated with motor commands. This interpretation aligns with the findings of \cite{mazzoni2007don}, who reported heightened sensitivity to movement costs but preserved movement accuracy in individuals with Parkinson’s disease relative to control subjects. While the nigrostriatal dopaminergic pathway (SNc to dorsolateral striatum) is well established as a key modulator of movement execution, the dopaminergic projections from the ventral tegmental area and, specifically, the mesolimbic dopaminergic pathway (VTA to medial striatum) are more strongly associated with learning processes. Multiple forms of learning—including skill learning and long-term motor adaptation—depend on these pathways (e.g., \citealp{krakauer2019motor,groenewegen2003basal,packard2002learning}; see also \cref{box:adaptations_learning}). Consistently, Parkinson's patients often show motor learning deficits \citep{olson2019motor,leow2012impaired,leow2013different}. Many of these learning processes require learning the rewarding nature of the body and world states. Such reward-state associations are precisely what a cost module must compute to assign costs to the estimated states. Collectively, current evidence is therefore compatible with the view that the basal ganglia are involved in the computation and/or storage of movement- and state-dependent costs \citep{shadmehr2008computational}.

\subsubsection{The controller module}
The controller generates motor commands that drive the system toward rewarding states while minimizing movement-related costs. This module is typically regarded as the core of the motor system, as it integrates information from all other modules. Early theories of motor control (e.g.,\cite{nelson1983physical,flash1985coordination,uno1989formation,harris1998signal,thoroughman2000learning}) typically conceptualized motor command generation as a sequential process comprising two submodules: motor planning and motor execution. In this framework, the planning stage computes an ideal trajectory that satisfies task demands, whereas the execution stage generates motor commands that drive the system along this predefined trajectory. A fundamental limitation of this control system is its lack of robustness to perturbations. For example, when unexpected disturbances displace the plant into states far from those anticipated during planning, adherence to the original trajectory may no longer be appropriate, necessitating an additional mechanism for replanning. In contrast, optimal feedback control (OFC) theory \citep{todorov2002optimal} posits that motor planning and motor command generation are inseparable. Under OFC, the controller learns the optimal time- and state-dependent policy $\bm{\pi}^*(t,\mathbf{x})$ that minimizes the expected cumulative cost over the task horizon $T_E$:
\begin{align}
J = \sum_{t=1}^{\mathrm{T_E}} \Bigl(
J_{\mathbf{x}}(t,\mathbf{x}(t)) + J_{\mathbf{u}}(t,\mathbf{u}(t))
\Bigr),
\end{align}
and generates motor commands online according to
\begin{align}
\mathbf{u}^*(t) = \bm{\pi}^*(t,\mathbf{x}(t)).
\end{align}

Owing to its central role in the generation of motor commands, motor cortex has naturally been regarded as a strong candidate for the neural substrate of the OFC theory controller. In the following sections, we examine the extent to which this hypothesis is supported by empirical evidence and consider its limitations \citep{shadmehr2008computational,scott2004optimal,haar2020revised}. 

One of the defining features of optimal feedback control (OFC) is its strong reliance on body and world state information. Specifically, within this framework, motor commands are generated in a state-dependent manner, where the state integrates external sensory feedback with internally generated predictions of sensory consequences provided by a forward model, which act as an internal feedback \citep{shadmehr2008computational}. The reliance of the motor system on feedback process is not an unprecedented feature. The motor system is, in fact, known to rely extensively on feedback processes implemented across multiple levels of the neuraxis. Reflex circuits provide rapid, automatic responses that serve to preserve the integrity of organism as a whole. For example, spinal stretch reflexes counteract sudden muscle lengthening, contributing to joint stability and posture while limiting excessive strain, whereas brainstem-mediated reflexes support fast defensive responses essential for survival. Importantly, these reflex pathways are not purely hard-wired but can be flexibly modulated by descending cortical and subcortical inputs, allowing reflex gains and response properties to be adjusted according to task demands and behavioral context \citep{shemmell2009differential}. Against this backdrop, it is natural to hypothesize that the motor system has evolved additional transcortical feedback loops to support more precise and flexible control of body state, operating through interactions between cortical and subcortical structures. 

If the control policy is state dependent, the neural substrate of the controller should exhibit corresponding signatures of this dependence. Consistent with this expectation, neurons in primary motor cortex (M1) are tuned not only to movement-related variables, such as reach direction, but also to sensory signals, including proprioceptive inputs associated with passive joint movements \citep{scott2004optimal}. During active movements, these sensory signals overlap with motor-related activity, indicating that M1 responses reflect integrated representations of the current state rather than purely motor commands \citep{scott1997reaching}. Similar state-dependent response properties have been reported in premotor cortex, where neural activity is modulated by visual information about surrounding objects relayed via posterior parietal cortex \citep{graziano1994coding,rizzolatti1988motor}, while also exhibiting clear planning- and movement-related modulation \citep{crammond2000prior}. 

Furthermore, the temporal profile of activity in primary motor cortex (M1) is consistent with the behavior expected of a reactive, state-dependent feedback controller. M1 neurons respond rapidly to sensory signals—for example, showing responses to limb perturbations within approximately 20 ms \citep{wolpaw1980correlations}—with this early activity likely driven by inputs from primary somatosensory cortex (S1). Later M1 responses, emerging after roughly 60 ms, appear to be strongly influenced by inputs from the cerebellum  \citep{alma9954958102101}, where responses to mechanical perturbations appear within about 20 ms \citep{scott2004optimal}. These later cortical signals contribute to long-latency muscle responses to perturbations, which account for intersegmental dynamics and support coordinated corrective actions.

Together, these findings are consistent with the view that motor and premotor cortices participate in state-dependent feedback control. Interestingly, this view is supported by a recent simulation study \citep{kalidindi2021rotational}, which demonstrated that rotational dynamics—a prominent feature of neural activity observed in motor cortex during the execution of goal-directed movements and often attributed to intrinsic recurrent connectivity—can instead emerge from strong sensory feedback signals reaching motor cortex and driving its dynamics. 

Another potential candidate is the cerebellum, which is widely believed to play a central role in correcting erroneous motor commands based on motor error (e.g., \citealp{herzfeld2018encoding,herzfeld2020principles}; see also \cref{sec:anatomy:cerebellum}). In this sense, cerebellar processing can be viewed as implementing a feedback mechanism that computes errors and generates corrective signals to counteract them. The pattern of inputs to and outputs from the cerebellum is consistent with this role. First, the cerebellum receives rich proprioceptive feedback from the spinal cord, as well as motor-related signals from the cortex that are often interpreted as efference copies of outgoing motor commands. Second, corrective signals can be conveyed via projections to the spinal cord through the brainstem, as well as through feedback pathways to cortical motor areas. Finally, as noted above, cerebellar responses to mechanical perturbations are remarkably rapid (on the order of 20 ms) and are thought to be relayed to motor cortex, where they contribute to shaping long-latency corrective responses. Together, these properties suggest that the cerebellum is also a plausible candidate for contributing to the implementation of the OFC controller.

\subsubsection{Limitations and open challenges}
Optimal feedback control (OFC) has provided a unifying and influential framework for understanding motor behavior. By integrating control, state estimation, forward modeling, and cost computation within a single normative theory, OFC accounts for a wide range of empirical findings and generates quantitative predictions that align closely with behavioral data. A notable example is the {\it minimal intervention principle} that the OFC controller implicitly follows by selectively correcting deviations along task-relevant dimensions while allowing variability to persist in task-irrelevant directions. This is an emergent feature of the OFC model that resembles human behavior, as observed across a variety of behavioral experiments \citep{scholz1999uncontrolled,scholz2000identifying}.

Despite these successes, several important limitations and open challenges remain. A first limitation concerns the level of abstraction of OFC relative to neuroanatomy. While OFC accurately captures the input–output behavior of the motor system and closely reproduces different aspects of human motor behavior, it remains unclear whether the internal computational modules posited by the framework correspond to those implemented by the brain. For example, evidence from rodents indicates that motor cortex is required for learning but not for the execution of well-learned motor behaviors \citep{kawai2015motor}, suggesting that execution may rely more heavily on subcortical structures. In this context, the basal ganglia and thalamus have been proposed to play a central role in action execution by facilitating the activation of appropriate brainstem and spinal motor circuits \citep{wolff2022distinct}. Moreover, additional structures in the spinal cord and cerebellum are known to provide corrective signals during ongoing movements \citep{sathyamurthy2020cerebellospinal}. In contrast, OFC concentrates control within a single controller module—associated with motor cortex—thereby abstracting away the distributed and hierarchical nature of control observed in biological motor systems. 

The second challenge concerns policy learning. OFC assumes that an optimal state-dependent policy is already available for the task at hand. In practice, however, learning such a policy is computationally demanding. For example, classical solutions require an iterative procedure that is repeated until convergence, which alternates between forward estimation of the Kalman gain and backward optimization of policy parameters \citep{todorov2002optimal}. How do brains find a solution to this optimization problem? If they do not optimize it directly, how do they learn to approximate the optimal policy? On which timescales? Which signals do they leverage? These remain open questions. 

A third, related limitation concerns the adaptation of the policy to deal with nonstationary dynamics. OFC typically assumes that task dynamics and costs are well defined, with a fixed forward model (i.e, $\hat{A}$, $\hat{B}$, $\hat{H}$) and cost structure (e.g., $\hat{S}(t)$, $\hat{M}(t)$ ). What happens when state dynamics change, e.g., when a perturbation is introduced that changes the mapping between motor commands and expected consequences? We know that when this happens, humans adapt their internal models accordingly, with a corresponding adjustment in their movements to counteract the perturbation. Within OFC, it is unclear whether such adaptation requires learning an entirely new policy or can be achieved by efficiently updating existing policies \citep{shadmehr1994adaptive}. Similarly, how the nervous system acquires and modifies task-dependent cost functions remains an open question.

Finally, OFC offers a limited account of motor planning and competition between action plans. Empirical evidence indicates that planning occurs even during self-paced movements \citep{lara2018conservation} and that multiple potential actions can be simultaneously represented within frontoparietal circuits \citep{cisek2010neural}. In standard OFC formulations, planning is subsumed into the learned policy, and the notion of competing plans is not explicitly represented. How OFC accommodates parallel action representations, and preprontal and basal ganglia mediated action selection remains insufficiently understood.

\subsection{Output modules for low-dimensional muscle coordination}
\label{sec:mod_motor_system:data_driven_output}
The human body possesses approximately 244 degrees of freedom (DOFs) \citep{morecki1984cybernetic}, actuated by roughly 630 skeletal muscles \citep{prilutsky2002optimization}. As a result, the musculoskeletal system is highly redundant at multiple levels. It exhibits {\it kinematic redundancy} allowing virtually all motor tasks to be performed using different joint configurations. It also exhibits {\it muscle redundancy}, because individual DOFs can be controlled by multiple muscles. Despite this extreme redundancy, the motor system achieves accurate and seemingly effortless control of movement. How has the nervous system rendered this high-dimensional control problem tractable? A long-standing proposal is that the motor system does not independently control each degree of freedom, but instead operates in a lower-dimensional control space \citep{bernstein1967bernstein}. In this view, movements are constructed through the flexible combination of a limited set of fixed motor modules, each corresponding to stereotyped submovements or spatiotemporal patterns of muscle activation.

Convincing evidence for the existence of such modules first emerged from stimulation studies in the frog spinal cord \citep{bizzi1991computations,giszter1993convergent,saltiel2001muscle}, later replicated in other animal models \citep{tresch1999responses,lemay2004modularity}. These studies showed that electrical or chemical stimulation of the spinal cord reliably evoked stereotyped patterns of muscle coactivation that depended on the site of stimulation. Each coactivation pattern generated a characteristic force field that drove the limb toward specific regions---or even discrete equilibrium points---of task space. Notably, the force fields tended to sum linearly: simultaneous stimulation of sites $A$ and $B$, which individually produced force fields $\mathbf{F}_A$ and $\mathbf{F}_B$, yielded a resultant force field well described by the vector sum $\mathbf{F_{AB}} = \mathbf{F}_A + \mathbf{F}_B$ \citep{mussa1994linear,lemay2001modulation}.

Subsequent work sought to identify such fixed coactivation patterns directly from muscle activity recorded during natural movements, typically using dimensionality reduction techniques such as nonnegative matrix factorization (NMF --- \cite{lee2000algorithms}). These studies generally reported that complex movements could be constructed through the flexible recruitment of a small number of fixed coactivation patterns, termed \emph{muscle synergies}, that coordinated activity across multiple muscles. These studies differed in their assumptions about the structure of individual synergies, variously modeling them as fixed patterns of muscle recruitment (spatial synergies), fixed temporal activation profiles (temporal synergies), or combined spatiotemporal patterns (spatiotemporal synergies). Nevertheless, they converged on the idea that synergies constitute a set of building blocks from which complex movements can be constructed. For example, \cite{ivanenko2004five} showed that five temporal activation patterns accounted for the activity of 25 muscles during human walking across five different speeds. Similarly, \cite{ting2005limited} demonstrated that four spatial muscle synergies explained the responses of 15 muscles during a challenging postural control task. Finally, \cite{d2006control} reported that five time-varying spatiotemporal synergies captured the muscle activity recorded from 19 muscles during reaching movements in both the frontal and sagittal planes. Subsequent studies clarified that muscle synergies may be task specific or shared across tasks \citep{d2005shared}, present at birth or emerging during development \citep{dominici2011locomotor}, and associated with the acquisition of specialized motor skills \citep{sawers2015long}. Moreover, simulation studies suggest that control strategies based on muscle synergies can reach acceptable levels of task performance more rapidly than strategies relying on independent control of individual muscles \citep{al2020effects,berg2024sar}. This has led to the proposal that synergistic control may provide a fast route to a functional, though not necessarily optimal, solution to motor tasks.

How low-dimensional factors extracted from muscle activity relate to underlying neural structures remains an open question. A large body of evidence implicates spinal circuits as a primary substrate for muscle synergies \citep{hart2010neural,levine2014identification,takei2017neural}, although contributions from the brainstem \citep{roh2011modules} and motor cortex \citep{overduin2012microstimulation,griffin2015corticomotoneuronal} have also been reported. Recordings from the frog spinal cord showed that activity of interneurons in the intermediate zone strongly correlates with muscle synergies extracted from electromyographic recordings \citep{hart2010neural}. In mice, optogenetic experiments identified a population of neurons in the medial deep dorsal horn that project to motoneurons, receive convergent inputs from corticospinal and sensory pathways, and drive reliable patterns of activation across multiple motoneuron pools spanning several spinal segments, thereby coordinating functionally related muscles \citep{levine2014identification}. Similarly, recordings from the primate cervical spinal cord revealed that premotor interneurons cluster into groups with spatial and temporal activity profiles consistent with muscle synergies observed during a precision grip task \citep{takei2017neural}. Complementary evidence from transection studies in frogs suggests that some motor modules are organized at the level of the brainstem \citep{roh2011modules}. While these findings point to subcortical circuits as a key locus of muscle synergies, there is also evidence for a role of cortical neurons. In particular, corticomotoneuronal cells in the motor cortex project to groups of functionally related muscles and may act as important nodes within synergistic control architectures \citep{griffin2015corticomotoneuronal}. This interpretation is consistent with microstimulation studies showing that muscle synergies extracted during natural behavior closely resemble the patterns of muscle activation elicited by stimulation of primary motor cortex \citep{overduin2012microstimulation}. Recent work has begun to clarify the complementary roles of spinal and cortical circuits in synergistic control \citep{takei2026primate}. During precision grip, spinal premotor interneurons accounted for the largest fraction of variance in muscle activity, consistent with a spinal encoding of core synergies, whereas corticomotoneuronal cells contributed additional inputs to smaller muscle groups, fine-tuning the expression of these spinally organized synergies.

As noted above, different modeling assumptions can be used to capture distinct invariances in muscle activation patterns. Although many formulations have been proposed---comprehensively reviewed elsewhere \citep{chiovetto2022toward,cheung2021approaches,singh2018systematic,tresch2006matrix}---in the following sections we focus on the two simplest and most representative classes of muscle synergy models, which capture modularity in the muscle and time domains, respectively. The problem addressed by muscle synergy models can be formalized as follows. The goal is to specify the instantaneous activation of $M$ muscles over $T$ time points, which is equivalent to defining the $MT$ entries of a matrix $\mathbf{Y} \in \mathbb{R}^{M \times T}$. A successful modularity model should reduce the dimensionality of this control problem by enabling muscle activity to be generated using substantially fewer parameters than the original $MT$ degrees of freedom.

\subsubsection{Spatial modules}

Spatial modularity models are based on the assumption that the motor system simplifies the control problem by relying on a set of \textbf{fixed spatial modules}
\begin{equation}
\mathbf{U} = (\mathbf{u}_1, \dots, \mathbf{u}_\mu),
\end{equation}
with $\mathbf{u}_i \in \mathbb{R}^{M}$ and $\mu < M$, which define $\mu$ spatial muscle recruitment patterns. According to this model, each column vector of $\mathbf{Y}$, denoted by $\mathbf{y}_{:,j} := (\mathbf{Y})_{:,j}$ and representing the activation pattern of all muscles at time $j$, can be generated by specifying a vector of \textbf{spatial activation coefficients} $\mathbf{a}_j \in \mathbb{R}^{\mu}$, according to
\begin{equation}
\mathbf{y}_{:,j} \approx \mathbf{U}\mathbf{a}_j .
\end{equation}

This strategy allows the motor system to specify only $\mu T$ spatial activation coefficients, rather than $MT$ individual muscle activation values. Accordingly, the full muscle activity matrix can be approximated as $\mathbf{Y} \approx \mathbf{U}\mathbf{A}$. In practice, this model is commonly fit to experimental muscle activity data by applying nonnegative matrix factorization (NMF) \citep{lee2000algorithms} to $\mathbf{Y}$ in order to estimate $\mathbf{U}$ and $\mathbf{A}$. The algorithm can be interpreted as identifying an optimal basis set $\mathbf{U}$ to linearly approximate the columns of $\mathbf{Y}$. Within this framework, each vector $\mathbf{u}_i$ is referred to as a spatial muscle synergy.

\subsubsection{Temporal modules}

Similarly, temporal modularity models are based on the assumption that the motor system simplifies the control problem by relying on a set of \textbf{fixed temporal modules}
\begin{equation}
\mathbf{V} = (\mathbf{v}_1, \dots, \mathbf{v}_\tau),
\end{equation}
with $\mathbf{v}_i \in \mathbb{R}^{T}$ and $\tau < T$, which define $\tau$ temporal muscle recruitment patterns. According to this model, each column vector of $\mathbf{Y}^\top$, denoted by $\mathbf{y}_{j,:} := (\mathbf{Y}^\top)_{:,j}$ and representing the activation pattern of muscle $j$ over time, can be generated by specifying a vector of \textbf{temporal activation coefficients} $\mathbf{b}_j \in \mathbb{R}^{\tau}$, such that
\begin{equation}
\mathbf{y}_{j,:} \approx \mathbf{V}\mathbf{b}_j .
\end{equation}

This strategy allows the motor system to specify only $M\tau$ temporal activation coefficients, rather than all $MT$ individual entries of $\mathbf{Y}$. In practice, this model is fit by applying nonnegative matrix factorization to $\mathbf{Y}^\top$, yielding $\mathbf{Y}^\top \approx \mathbf{V}\mathbf{B}$ and, equivalently, $\mathbf{Y} \approx \mathbf{B}^\top \mathbf{V}^\top$. Within this framework, each vector $\mathbf{v}_i$ is referred to as a temporal muscle synergy.

\subsubsection{Limitations and open challenges}

There is substantial evidence that muscle synergies reflect genuine features of the motor system, most notably from spinal cord stimulation experiments demonstrating the generation of force fields with convergence patterns that depend systematically on the site of stimulation. However, it remains less clear whether the synergies extracted from low-dimensional decompositions of muscle activity during voluntary behavior correspond to discrete neural modules, or instead emerge from task constraints---insofar as specific tasks admit only limited patterns of muscle activation---or from biomechanical constraints coupled with feedback mechanisms --- such as reflex pathways counteracting muscle stretch \citep{kutch2012challenges}. Although some observed synergies may indeed arise from such constraints (as testified by the presence of task-specific synergies \citep{d2005shared}), multiple lines of evidence argue against a purely task- or feedback-driven interpretation. 
For example, when task structure is altered by experimentally modifying the mapping between muscle activity and a virtually controlled object, participants more readily learn new activation patterns for existing synergies rather than entirely new synergies \citep{berger2013differences}. This finding suggests that synergies are constrained by underlying neural structures rather than being determined exclusively by task demands. Moreover, synergies have been shown to persist following deafferentation \citep{loeb1993effects}, indicating that they are not solely the product of sensory feedback. 

Muscle synergy models summarize high-dimensional EMG data using a low-dimensional set of invariant activation patterns, which may reflect neural structures leveraged by the motor system to simplify control. However, these models are primarily \emph{motor output} descriptions: they are largely concerned with how structures downstream of the cortex may be organized to reduce the dimensionality of the control problem faced by cortical areas. As such, they provide limited insight into the mechanisms by which the cortex selects an appropriate set of modules for a given task. In particular, synergy models do not specify how the nervous system determines, when confronted with a novel task, whether to recombine existing synergies using new activation patterns, to modify previously established synergies, or to acquire entirely new ones. In general, synergy models do not account for the mechanisms and timescales of the learning processes necessary to perform novel tasks. Moreover, even under the assumption that all relevant modules have already been learned, muscle synergy models are fundamentally static and do not specify the dynamics underlying the generation of the temporal signals present in the model. For example, they do not address how supraspinal networks generate the activation profiles that modulate spatial synergies over time, $(\mathbf{A})_{i,:} \in \mathbb{R}^{T} $, nor how spinal circuits generate the temporal structure captured by temporal synergies, $\mathbf{v}_i \in \mathbb{R}^{T}$.

An additional challenge emerges when considering how activation patterns must be specified to recruit synergies across different movements. In this setting, the control problem involves specifying a set of muscle activation matrices $\mathbf{Y} \in \mathbb{R}^{M \times T \times R}$, where $R$ denotes the number of distinct movements. From this perspective, classical muscle synergy models do not scale favorably with the number of movements. In spatial synergy models, $\mu$ spatial modules are fixed across movements and over time, but each movement requires specifying a distinct matrix of spatial activation coefficients of dimensionality $\mu \times T$. As a result, a total of $\mu T R$ coefficients must be specified. Similarly, in temporal synergy models, $\tau$ temporal modules are fixed across movements and across muscles, but each movement requires specifying a matrix of temporal activation coefficients of dimensionality $M \times \tau$, leading to a total of $M \tau R$ coefficients. Recent modeling approaches aim to address this limitation by jointly capturing modularity in both the spatial and temporal domains. By adopting alternative modeling assumptions, these approaches further reduce the dimensionality of the control problem, decreasing the number of required activation coefficients to $\mu \tau R$ \cite{delis2014unifying}, and in some cases to as few as $S R$ \citep{salatiellohierarchical}, where $S$ denotes the number of coupled spatiotemporal synergies.

\subsection{Dynamical modules for smooth pattern generation}
\label{sec:mod_motor_system:data_driven_brain}
In the following sections, we examine the \textit{dynamical systems view} of motor control \citep{churchland2024preparatory,vyas2020computation,churchland2012neural} and the dynamical modules that have emerged from this framework. This perspective arose in response to empirical observations of motor cortical activity that proved difficult to reconcile with classical representational models. Such models sought to directly relate the neural activity of a population of neurons, $\mathbf{r}(t) \in \mathbb{R}^N$, to a set of behaviorally relevant movement parameters, $\mathbf{p}(t) \in \mathbb{R}^M$, such as movement direction \citep{georgopoulos1982relations} or force \citep{evarts1968relation}. In their simplest form, these models posit a linear relationship,
\begin{equation}
\mathbf{r}(t) = A \mathbf{p}(t) + \mathbf{b},
\end{equation}
where $A \in \mathbb{R}^{N \times M}$ specifies the tuning of each neuron to the movement parameters and $\mathbf{b} \in \mathbb{R}^N$ denotes baseline firing rates.

A canonical example is the cosine-tuning model for two-dimensional reaching movements, in which velocity is taken as the relevant movement parameter. In this case, the firing rate of neuron $i$ is modeled as
\begin{equation}
r_i(t) = b_i + \mathbf{a}_i^\top \mathbf{v}(t),
\end{equation}
where $\mathbf{v}(t) \in \mathbb{R}^2$ is the hand velocity and $\mathbf{a}_i^\top \in \mathbb{R}^2$, the i-th row of A, corresponds to the i-th neuron's preferred-direction vector. This expression can be rewritten as
\begin{equation}
r_i(t) = b_i + \|\mathbf{a}_i^\top\|\,\|\mathbf{v}(t)\| \cos\!\big(\alpha_i - \beta(t)\big),
\end{equation}
where $\alpha_i$ denotes the preferred direction of neuron $i$ and $\beta(t)$ the instantaneous movement direction. Thus, under the cosine-tuning model, neuron $i$ fires most strongly when the instantaneous movement direction is aligned with its preferred direction.

Several major challenges for representational models such as cosine tuning, are apparent. First, many neurons have been found to be tuned to a wide range of both high- and low-level movement parameters \citep{kakei1999muscle,omrani2017perspectives}. Second, and more difficult to explain, neuronal tuning often varies with context and movement phase \citep{churchland2007temporal}. For example, tuning properties can change markedly between movement preparation and execution \citep{churchland2010cortical}. Third, natural movement parameters are inherently correlated, such that a neuron tuned to one variable will often appear sensitive to other movement variables at different time lags \citep{reimer2009problem}. Finally, representational theories typically do not specify how muscle activity is generated from neural population activity $\mathbf{r}(t)$. This transformation—from cortical activity to muscle commands—is a computationally demanding problem and should be central to any comprehensive theory of motor control. Together, these observations indicate that neural activity cannot be straightforwardly explained as reflecting fixed preferences for specific movement parameters, and suggest that an alternative theoretical framework may be better suited to capture the observed activation patterns and their underlying computational structure \citep{omrani2017perspectives,churchland2012neural,scott2008,Fetz_1992}. 

Unlike representational models, the \textit{dynamical systems view} of motor control \citep{churchland2024preparatory,vyas2020computation,churchland2012neural} conceptualizes motor cortex as a pattern generator that produces rich intrinsic dynamics—akin to a basis set or a reservoir in reservoir computing \citep{jaeger2001echo,maass2002real}—that downstream circuits can flexibly read out to generate muscle activity. In this framework, population activity evolves according to recurrent dynamics modulated by external inputs,
\begin{equation}
\dot{\mathbf{r}}(t) = \mathbf{f}\big(\mathbf{r}(t)\big) + \mathbf{u}(t),
\end{equation}
where $\dot{\mathbf{r}}(t)$ denotes the time derivative of the firing-rate vector $\mathbf{r}(t)$ and $\mathbf{u}(t)$ represents external inputs, for example, from upstream cortical areas. Muscle activity $\mathbf{m}(t) \in \mathbb{R}^E$ is then generated by a readout of population activity,
\begin{equation}
\mathbf{m}(t) = \mathbf{g}\big(\mathbf{r}(t)\big),
\end{equation}
with $\mathbf{g}(\cdot)$ often modeled as a linear mapping.

For this scheme to be viable, additional processes are required to prepare movements without prematurely activating muscles, control movement timing, and ensure robustness to noise and contextual variability. These processes are thought to operate largely within output-null subspaces of population activity, such that they do not directly influence muscle output. From the dynamical systems perspective, these processes can be interpreted as dedicated dynamical modules with distinct computational roles. Below, we discuss the principal modules for which experimental evidence has been reported. Note that these computational modules are inferred by the observation and subsequent characterization of consistent and structured activation patterns within low-dimensional subspaces of neural population dynamics across different movement phases. The task most commonly used to cleanly separate these phases is the instructed delayed-reaching task, which enables a clear distinction between movement preparation—spanning the interval from target presentation to the go cue—and movement execution, which begins after the go cue.

\subsubsection{The motor preparation module}

The motor preparation module unfolds during the preparatory period following target presentation, which specifies the movement to be executed when the go cue appears. During this phase, two robust features of neural population activity are observed. First, trial-to-trial variability is markedly reduced, a phenomenon also observed in other cortical areas following stimulus presentation \citep{churchland2010stimulus}. Second, population activity converges toward condition-specific regions of state space. Importantly, such preparatory dynamics are observed even in the absence of an experimenter-imposed delay \citep{lara2018conservation}, suggesting that they play a mechanistic role rather than reflecting a mere epiphenomenon. Additionally, the resulting preparatory states closely resemble condition-specific fixed points, as demonstrated by optogenetic perturbation experiments \citep{inagaki2019discrete}. Optogenetic perturbations delivered during preparation transiently displace population activity, but the movement that is ultimately executed is determined by the state-space region to which activity subsequently converges: return to the original preparatory region yields the correct movement, whereas convergence to a different region results in execution of the corresponding incorrect movement \citep{inagaki2019discrete}. These findings support the interpretation of motor preparation as a process that sets the system’s initial conditions for subsequent movement execution \citep{churchland2010cortical}. Consistent with this view, recurrent neural networks trained to generate muscle activity from condition-dependent initial states (under constraints favoring smooth dynamics) exhibit activation patterns that closely resemble those observed in motor cortex \citep{hennequin2014optimal,sussillo2015neural}. This provides computational support for the plausibility of preparatory activity functioning primarily to initialize the dynamical system.

A key question is how preparatory activity avoids prematurely driving muscle activity. Evidence suggests that this is achieved by confining population activity during preparation largely to output-null subspaces of the neural state space \citep{kaufman2014cortical}. During movement execution, neural activity transitions into output-potent subspaces that directly influence muscle activity. Formally, neural population activity $\mathbf{r}(t)$ can be decomposed as
\begin{equation}
\mathbf{r}(t) = \mathbf{r}_{\mathrm{prep}}(t) + \mathbf{r}_{\mathrm{exec}}(t),
\end{equation}
where $\mathbf{r}_{\mathrm{prep}}(t)$ lies in the output-null subspace and $\mathbf{r}_{\mathrm{exec}}(t)$ lies in the output-potent subspace. Under a readout $\mathbf{g}()$ mapping neural activity to muscle activity,
\begin{equation}
\mathbf{g}\big(\mathbf{r}_{\mathrm{prep}}(t)\big) \approx \mathbf{0}, 
\qquad
\mathbf{g}\big(\mathbf{r}_{\mathrm{exec}}(t)\big) \neq \mathbf{0}.
\end{equation}
During movement preparation, $\mathbf{r}_{\mathrm{prep}}(t) \neq \mathbf{0}$ while $\mathbf{r}_{\mathrm{exec}}(t) \approx \mathbf{0}$; during movement execution, the pattern is reversed such that $\mathbf{r}_{\mathrm{exec}}(t)$ dominates the population response.

\subsubsection{The timing module}
The timing module is engaged following presentation of the go cue, which signals that the prepared movement should be initiated. During this phase, a prominent feature of neural population dynamics is a large, condition-invariant translation of the population state \citep{kaufman2016largest}. This translation is mediated by a condition-invariant signal (CIS) that accounts for a substantial fraction of the population variance (more than half) and shifts neural activity from a preparatory region of state space, characterized by attractor-like dynamics, into an execution region with dynamics more suitable for pattern generation. Consistent with this interpretation, movement-related dynamics begin to unfold toward the end of the CIS. Importantly, the timing of the CIS is strongly predictive of movement onset rather than of go-cue presentation, suggesting that it reflects internally generated timing signals rather than a direct sensory response. This observation has led to the hypothesis that the CIS may originate from brain areas involved in temporal control, such as the cerebellum \citep{meyer1977cerebellar} or the basal ganglia \citep{hauber1998involvement,romo1992role}. Supporting the plausibility of this mechanism, recurrent neural network models can generate appropriately timed, EMG-like muscle activity with motor-cortex-like population dynamics when driven by a condition-invariant go signal \citep{hennequin2014optimal,sussillo2015neural}.

\subsubsection{The motor execution module}

The motor execution module becomes active shortly before movement onset and unfolds throughout movement execution. During this phase, a dominant feature of neural population dynamics is the presence of rotational structure in state space, with condition-dependent amplitude and phase but largely condition-invariant frequency \citep{churchland2012neural}. This rotational structure was revealed by observing that, during movement execution, population activity is well approximated by a linear time-invariant dynamical system
\begin{equation}
\dot{\mathbf{r}}(t) = A \mathbf{r}(t),
\end{equation}
with a skew-symmetric state matrix
\begin{equation}
A = -A^\top.
\end{equation}

These rotational dynamics, typically in the range of 1.5--3~Hz \citep{lara2018different}, are prominent at the population level—for example, the most dominant rotational plane can account for approximately 30\% of the total variance in neural activity—yet are not directly evident in electromyographic (EMG) signals. Nevertheless, EMG activity can be accurately reconstructed from a small number of such neural oscillatory components with condition-dependent amplitude and phase \citep{churchland2012neural}. This observation suggests that rotational dynamics provide an efficient basis for generating muscle activity. This interpretation is further supported by studies exploring the ability of recurrent neural network models to generate EMG-like outputs \citep{sussillo2015neural,salatiello2020recurrent}. When these models are trained under constraints favoring simple or smooth dynamics, they naturally develop rotational population activity resembling that observed in motor cortex; by contrast, such rotational structure is absent when these constraints are removed \citep{sussillo2015neural}. When these models are trained to generate EMG-like patterns while exhibiting motor cortex-like rotational dynamics, they generalize significantly better to untrained EMG patterns, with generalization performance improving as similarity to more components of the rotational dynamics is enforced \citep{salatiello2020recurrent}.

Critically, this rotational structure is not ubiquitous across motor-related cortical areas. For example, it has been consistently observed in primary motor cortex (M1) and dorsal premotor cortex (PMd) \citep{churchland2012neural}, but not in supplementary motor area (SMA) \citep{lara2018different}. This dissociation suggests that rotational dynamics may be particularly suited to areas such as M1 and PMd, which are more directly involved in pattern generation and have closer access to subcortical motor circuits. Conversely, different dynamical regimes may be more appropriate for areas that are thought to play distinct functional roles. For example, activity in SMA exhibits helical rather than purely circular trajectories: during the execution of repeated, identical movements, population activity progressively translates through state space, occupying different regions on successive repetitions \citep{russo2020neural}. Such dynamics are well suited for encoding contextual information, enabling movements to be terminated once the required number of repetitions has been completed.

A recent study provided further insight into the functional role of rotational dynamics in motor cortex. Using population-level analyses, \citet{russo2018motor} suggested that such dynamics enable motor cortex to represent muscle activity patterns while keeping state-space dynamics relatively untangled, such that nearby states evolve with similar flow fields. To quantify this property, they introduced a measure of dynamical tangling,
\begin{equation}
Q(t) = \max_{t'}
\frac{\left\lVert \dot{\mathbf{r}}(t) - \dot{\mathbf{r}}(t') \right\rVert^2}
{\left\lVert \mathbf{r}(t) - \mathbf{r}(t') \right\rVert^2 + \varepsilon},
\end{equation}
where $\lVert \cdot \rVert$ denotes the Euclidean norm and $\varepsilon$ is a small constant included to prevent division by zero.

Using this metric, the authors showed that tangling is high for electromyographic (EMG) signals—reflecting the fact that similar muscle activation patterns can be followed by divergent future activity across movements—as well as in sensory regions such as S1 and V1, which are strongly driven by external inputs. In contrast, tangling is low in motor cortex, which is directly involved in movement generation and must therefore operate in a noise-robust regime. Low tangling ensures that small perturbations to population activity do not lead to qualitatively different dynamical trajectories and, consequently, do not result in unintended movements.

Moreover, they demonstrated that directly optimizing neural population activity to simultaneously linearly encode muscle activity and minimize tangling yields solutions that naturally exhibit rotational structure, closely resembling experimentally observed motor cortical dynamics. These results led to the conclusion that allowing efficient representations of muscle signals while minimizing dynamical tangling is a key computational principle of motor cortex. Importantly, the authors emphasized that although rotational dynamics emerge in the tasks studied, including reaching and cycling tasks, the specific dynamical structure that minimizes tangling may depend on task demands and need not always be rotational.

Finally, although early interpretations proposed that these rotational dynamics arise autonomously from intrinsic recurrent connectivity within motor cortex, more recent work indicates that external inputs play a critical role in sustaining them. In particular, thalamic input has been shown to be necessary for maintaining appropriate motor cortical dynamics during dexterous arm movements \citep{sauerbrei2020cortical}. Consistent with the idea that external inputs may drive these oscillations, subsequent work provided evidence that sensory feedback to motor cortex can contribute to rotational dynamics \citep{kalidindi2021rotational}. Using simulations of arm movements driven by a network model, this study showed that sensory feedback signals themselves exhibit rotational structure and that these dynamics are sufficient to induce rotations in the controller network, even in the absence of recurrent connectivity. Moreover, rotational dynamics were observed in somatosensory (S1, area 2) and parietal (area 5) cortices during delayed-reaching tasks in monkeys, supporting the idea that such dynamics may originate upstream of motor cortex. 

\subsubsection{Limitations and open challenges}

Decades of work have shown that purely representational models provide limited insight into the computations performed by motor cortex. Such models struggle to account for phenomena such as simultaneous tuning to multiple high- and low-level movement parameters, as well as context- and phase-dependent changes in tuning between movement preparation and execution. Moreover, given that motor cortex is the cortical area with the most extensive downstream projections to subcortical motor structures, it is intuitively plausible that its primary role is to generate appropriate population activity patterns to drive downstream circuits, rather than to explicitly encode movement parameters that must later be transformed into muscle commands. From this perspective, the dynamical systems view offers a more suitable conceptual framework.

The dynamical systems approach has indeed revealed several prominent motifs in neural population dynamics—which can be interpreted as dynamical modules—with plausible links to computational function. These include convergence to condition-specific regions of state space during movement preparation to set initial conditions, condition-invariant translations associated with movement timing, and rotational dynamics during execution that support the generation of muscle activity while maintaining smooth, untangled trajectories. However, a major unresolved issue is whether these dynamical features arise primarily from intrinsic cortical dynamics or are driven by external inputs. If the latter, the sources of these inputs and their relationship to sensory cues and task structure remain poorly understood. Furthermore, the apparent engagement of distinct dynamical regimes across different movement phases raises the question of whether a single, unified dynamical system underlies motor control. More generally, while the dynamical systems framework is effective at describing observed population activity, it has yet to yield compact, interpretable equations governing the emergence of these motifs. By contrast, optimal feedback control theory (\cref{sec:mod_motor_system:engineering}) provides such equations at the behavioral level, but makes relatively weak predictions about the specific dynamical features that should be observed in neural population activity.

Finally, much of the dynamical systems view of motor control is strongly motor-cortex-centric. Although motor cortex is clearly critical for motor learning and for fine, goal-directed movements—particularly in primates—its necessity for motor behavior varies across tasks and species. In many animals, including rodents, a wide range of movements can be executed in the absence of cortex, relying instead on brainstem and spinal circuits that are themselves capable of generating rich motor patterns. This raises the possibility that motor cortical activity patterns may not always directly control muscles, but instead provide higher-level signals that coordinate downstream motor circuits. Consistent with this idea, different dynamical regimes have been observed in motor cortex during different classes of movements. For example, low tangling has been reported during reaching, cycling, and locomotion \citep{russo2018motor}, but not during grasping movements \citep{suresh2020neural}. This dissociation has been interpreted as reflecting continuous inputs from premotor and sensory areas during grasping \citep{suresh2020neural,churchland2024preparatory}, but may also indicate a shift in motor cortical function—from high-level coordination of downstream circuits to more direct control of individual muscles. In contrast to motor-cortex–centric theories of motor control, such as the dynamical systems view, muscle synergy theories (\cref{sec:mod_motor_system:data_driven_output}) come closer to explaining how subcortical modules might be leveraged to facilitate motor control. They explicitly aim to capture invariances at the output level of the motor system, which likely reflect reliance on a repertoire of subcortical modules. Nevertheless, like optimal feedback control theory, they provide limited predictions regarding the dynamical structure of cortical population activity.

\section{Conclusion}
\label{sec:conclusion}
Controlling the human body is a challenging, high-dimensional, and dynamic problem. The nervous system must continuously transform noisy and delayed sensory measurements into motor commands that drive a redundant musculoskeletal plant through a changing world, often in the absence of explicit or stable rewards. Yet human movements are typically fast, accurate, and remarkably robust—all while operating under severe energetic constraints. A wide range of complementary principles has been proposed to explain how the brain achieves such computational and energetic efficiency. For instance, neural circuits exploit {\it sparse} \citep{olshausen2004sparse} and {\it distributed} \citep{steinmetz2019distributed} representations to encode sensory information efficiently and robustly, facilitating downstream processing and integration \citep{rigotti2013importance}. They build and leverage internal {\it world models} to continuously {\it predict} sensory input, allowing attention to be selectively allocated to the most unexpected and informative signals \citep{aizenbud2025neural}. These predictions also support Bayesian-optimal state estimation \citep{kording2004bayesian} and action selection \citep{KORDING2006319}. In addition, the brain leverages multiple learning systems \citep{squire1992declarative} to enable continual learning \citep{kudithipudi2022biological} of increasingly more abstract concepts \citep{behrens2018cognitive}. Another key computational principle supported by strong empirical and theoretical evidence is modularity \citep{bullmore2012economy,salatiello2025modularity}. A central theme of this review is that \emph{modularity} constitutes a fundamental organizational principle of brain structure and function. In the first part of the review, we surveyed converging evidence that modular organization is a general property of nervous systems, drawing on lesion studies, parcellation approaches, and network analyses. We then discussed candidate mechanisms that may favor the emergence and evolutionary selection of modular architectures, emphasizing the intrinsic robustness, flexibility, energetic efficiency, and computational advantages conferred by modular organization.

In the second part of the review, we focused on the motor system, examining modularity from both neuroanatomical and computational perspectives. We first reviewed the major \emph{anatomical modules} of the motor system, ranging from subcortical structures closely tied to motor output to cortical regions that are strongly shaped by sensory inputs. We then discussed how these modules may interact during different movement phases. Next, we considered three major computational theories of motor control and showed that each is, at its core, modular. In optimal feedback control (OFC) theory, motor behavior emerges from the interaction of distinct \emph{computational modules} responsible for system identification, state estimation, cost computation, and control policy generation. This decomposition is explicit and normative: each module has well-defined inputs, outputs, and functional roles, and the theory specifies how their interactions give rise to movement. In motor primitive and muscle synergy frameworks, \emph{motor output modules} encode useful spatiotemporal coordination patterns that reduce the effective dimensionality of the control problem and mitigate musculoskeletal redundancy. Finally, dynamical systems perspectives highlight the role of \emph{dynamical modules} that generate cortical population dynamics, preparing movements within output-null subspaces and encoding muscle activity patterns through smooth flow fields. Such dynamics promote robustness and allow downstream subcortical centers to read out appropriate muscle patterns with minimal additional computation. While these frameworks emphasize different computational objects---policies and costs (OFC), coordination patterns (primitives/synergies), or population dynamics (dynamical systems)---their common ground is striking: each posits that complex movements emerge from the interaction of specialized components whose composition reduces computational complexity and improves robustness. This convergence suggests that modularity is not a narrow hypothesis about a particular circuit but a multi-level organizing principle of the motor system.

At the same time, each framework faces open challenges that point to a shared agenda for the field. OFC theory is particularly compelling as a normative account: it specifies what each computational component should do and why. However, mapping these components onto neural substrates remains a major challenge. It is plausible that such computational modules are implemented in the brain through distributed, recurrent interactions among multiple anatomical modules. A key open question, therefore, is how the computational components of OFC can be related to neural circuitry. One promising approach is to analyze the activation dynamics of recurrent neural networks (RNNs) trained to perform tasks analogous to those assigned to OFC modules, while embedding these networks in realistic sensorimotor loops coupled to physically grounded plants. Such models allow one to ask whether the resulting dynamics resemble those observed in specific motor areas and whether particular computational roles can be tentatively associated with identifiable neural substrates. Following a similar approach, \cite{kalidindi2021rotational} trained an RNN to control a biomechanical arm model using realistic sensory feedback. They found that the activity of the network’s feedback controller exhibited dynamical features resembling those observed in primary motor cortex (M1), providing support for OFC-based interpretations of motor control and offering a tentative mapping between an OFC module and a cortical area. Importantly, their results also suggested that the rotational dynamics observed in M1 may be driven in large part by structured sensory input rather than by intrinsic recurrent connectivity alone, thereby informing and constraining dynamical systems accounts of motor control.

Muscle synergy frameworks, by contrast, are grounded in substantial electrophysiological, pharmacological, and behavioral evidence and offer an intuitive solution to the problem of musculoskeletal redundancy. However, such models remain largely \emph{static} descriptions of motor output structures. In particular, they often do not specify how upstream, cortical motor areas orchestrate the subcortical synergy-encoding modules, how temporal activity patterns emerge from circuit dynamics, or how sensory feedback shapes the recruitment or activation dynamics of the synergies. As a result, it remains difficult to identify the neural substrates responsible for coordination, timing, and pattern generation within these frameworks. Relatedly, these approaches leave open a fundamental question: how does a motor system organized around synergies learn new movements? How does it determine whether a novel task can be accomplished by reweighting or fine-tuning existing primitives, or whether the creation of new synergies is required? Although a comprehensive theory of primitive-based motor control and learning is still lacking, several studies provide encouraging evidence that controllers built from motor primitives can effectively control complex plants \citep{berniker2009simplified} and achieve faster convergence to high-performance solutions than non-modular alternatives \citep{al2020effects}. An important next step is to integrate these computational insights with neurobiologically plausible learning rules and circuit-level implementations, enabling explicit links between primitive-based control, learning, and identifiable neural substrates.

Dynamical systems perspectives have been transformative in reframing motor cortex from a static representational map into a dynamical system responsible for pattern generation. However, current formulations often resemble a collection of dynamical features observed across different movement phases and cortical areas, rather than a fully unified theory. A key next step is to define a principled framework that explains how these features arise from interacting \emph{dynamical modules}, how these modules coordinate across movement phases, and how their interactions are constrained by the mechanics of the body and the statistics of sensory feedback. Further progress will also require integrating the contributions of other components of the motor hierarchy, including the brainstem, spinal cord, cerebellum, basal ganglia, and thalamus. Recent work by \cite{logiaco2021thalamic} represents a promising step in this direction, proposing a unifying model of how basal ganglia, thalamus, and motor cortex interact to generate and flexibly recombine motor motifs, reminiscent of motor primitives. In this model, the basal ganglia select which motif should be prepared and determine the timing of its execution by selectively disinhibiting specific thalamocortical loops. Engagement of these loops then gives rise to motif-specific dynamics, which include an obligatory preparation phase followed by rotational dynamics during execution—features consistent with population activity patterns observed in motor cortex. Interfacing multi-area models of this kind with embodied plants and realistic sensory feedback provides a natural path toward a more comprehensive theory of modular motor control.

Taken together, the path forward is clear: progress will likely require \emph{closing the loop} between theory, multi-area recordings, and embodied simulations that model plants, physical environments, and sensory feedback. From the experimental side, a major advance will come from large-scale, multi-area recordings obtained during the execution of quantitatively tracked natural behaviors over extended timescales \citep{steinmetz2021neuropixels,markowitz2018striatum,dhawale2017automated}. On the modeling side, increasingly realistic musculoskeletal simulators and physics-based environments will be essential not only for generating realistic sensory feedback and linking control signals to task-level kinematics, but also for rigorously testing the hypothesized benefits of modular control frameworks—such as robustness, reuse, and accelerated learning—when controlling high-dimensional plants. The most informative theories will be those that (i) specify explicit computations and interactions among modules within a coherent framework, (ii) generate quantitative predictions for neural dynamics across multiple brain areas and timescales, and (iii) generalize to novel trained movements or distinct phases of natural behavior. In summary, despite differences in emphasis and level of abstraction, computational theories of motor control converge on a common claim: modularity is a fundamental organizational principle of the motor system, which provides clear computational benefits. The next generation of work should aim to explain how modules are selected, composed, and adapted to produce flexible movement in a physical world—and how these processes are implemented by interacting cortical and subcortical circuits operating in closed loop with the body.

\bibliographystyle{unsrtnat}
\bibliography{sections/_references}  %%% Uncomment this line and comment out the ``thebibliography'' section below to use the external .bib file (using bibtex).

\end{document}